\documentclass[fleqn,usenatbib]{mnras}
\usepackage{newtxtext,newtxmath}
\usepackage[T1]{fontenc}

\DeclareRobustCommand{\VAN}[3]{#2}
\let\VANthebibliography\thebibliography
\def\thebibliography{\DeclareRobustCommand{\VAN}[3]{##3}\VANthebibliography}

\usepackage{tablefootnote}
\usepackage{threeparttable}
\usepackage{subcaption}
\usepackage{booktabs}
\usepackage{orcidlink}
\usepackage[export]{adjustbox}
\usepackage{graphicx}	% Including figure files
\usepackage{amsmath}	% Advanced maths commands
\usepackage{pdflscape}

\title[Mass loss and clumping in the Arches Cluster]{ALMA measurements of mass loss and wind clumping in the massive stars of the Arches cluster}

\author[J. P. Perry et al.]{
James P. Perry,$^{1}$\thanks{E-mail: james.perry.21@ucl.ac.uk}\orcidlink{0000-0002-2090-6892}
Raman K. Prinja,$^{1}$\orcidlink{0000-0002-5251-3743}
Danielle M. Fenech$^{2}$\orcidlink{0000-0001-8443-1649}
and Francisco Najarro$^{3}$\orcidlink{0000-0002-9124-0039}
\\
$^{1}$Department of Physics and Astronomy, University College London, WC1E 6BT London, UK\\
$^{2}$SKAO, Jodrell Bank, Lower Withington, Macclesfield, UK\\
$^{3}$Centro de Astrobiolog\'ia, CSIC-INTA, Ctra de Torrej\'on a Ajalvir km 4, 28850 Torrej\'on de Ardoz, Madrid, Spain
}

\date{Accepted XXX. Received YYY; in original form ZZZ}

\pubyear{\the\year{}}

\begin{document}
\label{firstpage}
\pagerange{\pageref{firstpage}--\pageref{lastpage}}
\maketitle

\begin{abstract}
We present the first Atacama Large Millimeter/submillimeter Array (ALMA) Band 3 (100 GHz) and Band 6 (243 GHz) continuum observations of the Arches cluster, one of the youngest and most massive stellar clusters in the Milky Way. We detect and characterise millimetre emission from 23 massive stars, including WN7-9h Wolf-Rayet stars, O-type supergiants and hypergiants. By combining our ALMA measurements with archival Very Large Array data spanning 5--22.5 GHz, we derive broadband radio-millimetre spectral indices and investigate the radial structure of stellar winds through frequency-dependent clumping diagnostics. The majority of Wolf-Rayet stars exhibit spectral indices clustered around $\alpha \approx 0.7-0.8$, consistent with predominantly thermal free-free emission from dense, partially optically thick winds. In contrast, several O-type stars show flat or negative broadband spectral indices, indicative of non-thermal synchrotron emission likely associated with colliding-wind binaries. Using millimetre flux densities, we derive clumping-scaled mass-loss rates spanning $\log(\dot{M}/\mathrm{M}_{\odot}\,\text{yr}^{-1})\approx-4.1$ to $-4.9$ for the WN stars and $-4.9$ to $-5.4$ for the O super-/hypergiants, consistent with expectations for luminous massive stars in the Galactic Centre environment. We find significant evidence of structured wind clumping at millimetre wavelengths that generally decreases with increasing radius, supporting structured wind models with strong inner-wind inhomogeneities. These results demonstrate the power of combined radio-millimetre observations for constraining mass-loss and wind structure in massive stars, and provide new insight into stellar feedback in extreme cluster environments.
\end{abstract}


%%%%%%%%%%%%%%%%%%%%%%%%%%%%%%%%%%%%%%%%%%%%%%%%%%

%%%%%%%%%%%%%%%%% BODY OF PAPER %%%%%%%%%%%%%%%%%%
\section{Introduction}
Throughout their lifetimes, massive stars ($M>8\,\mathrm{M}_{\odot}$) drive galactic evolution. They shape the interstellar medium (ISM) by injecting both mechanical and ionising radiation via winds, supernovae and gamma-ray bursts. They are also the principal drivers of galactic chemical evolution through the creation and deposition of heavy elements. Observationally, massive stars dominate emission at UV, X-ray, infrared and radio wavelengths. Yet compared to their lower-mass counterparts, massive stars comprise only a small fraction of the Galactic stellar population. Given that they also have significantly shorter lifetimes and are predominantly found at large distances, massive stars are much less well understood. 

A critical factor in the evolution of massive stars (along with rotation and magnetic fields) is mass loss, primarily through radiation-driven stellar winds \citep{ekstrom12,josiek24}. These winds are especially strong in hot, luminous O-type and Wolf-Rayet (WR) stars, where the momentum transfer from photons to ions in the stellar atmosphere drives continuous outflows. Empirical and theoretical studies have established typical mass-loss rates for O-type stars in the range of $\dot{M} \sim 10^{-6}$ to $10^{-5}\,\mathrm{M}_{\odot}\,\text{yr}^{-1}$, depending on luminosity and metallicity \citep{vink01}. For WR stars, which represent evolved massive stars that have shed the majority, or all, of their hydrogen envelopes, mass-loss rates can reach $\dot{M} \sim 10^{-5}$ to $10^{-4}\,\mathrm{M}_{\odot}\,\text{yr}^{-1}$ \citep{nugis00}.

The cumulative effects of such mass loss can significantly alter the evolutionary trajectories of massive stars, causing them to lose substantial mass through stellar winds and eruptive events during main-sequence and post-main-sequence evolution. The stripping exposes deeper layers, changes the surface composition, and affects the star’s position in the Hertzsprung-Russell (HR) diagram. This mass loss ultimately influences the final fate of the star -- determining whether it ends as a neutron star, black hole, or collapses directly without a supernova \citep{groh13}.

In massive star clusters like the Arches, where the stellar metallicity is near-solar or slightly supersolar \citep{najarro04}, mass-loss rates are expected to be enhanced due to the metallicity dependence of radiatively driven winds \citep[$\dot{M} \propto Z^{0.7}$;][]{vink01}. Combined with the high stellar densities that may promote dynamical interactions or binary mass transfer, the mass-loss histories of Arches cluster stars are likely more complex than in lower-density environments.

First observed by \citet{nagata95} and \citet{cotera96}, the Arches cluster is one of the youngest \citep[$\sim$2-4 Myr; e.g.][]{figer99,figer02,najarro04,martins08,schneider14} and most massive \citep[$(2-6)\times 10^{4}\,\mathrm{M}_{\odot}$; e.g.][]{kim00,clarkson12} clusters known in the Milky Way. It is located at a distance of approximately 30 pc (in projection) from the Galactic Centre (GC), placing it within the 100 pc radius of the Central Molecular Zone (CMZ). As a result of the high extinction and crowding near the GC, observations have been limited to the near-infrared (NIR), radio and X-ray \citep[see e.g.][]{figer02,dong11,cano21,gallego21,capelli11}. 

These studies have identified more than 100 O-stars and many WNLh stars. In particular, \citet{martins08} sampled 28 of the brightest cluster members. By combining \textit{K}-band VLT/SINFONI spectra with atmospheric models computed with CMFGEN, they derived fundamental parameters of the sample, including effective temperatures, luminosities, stellar abundances, mass-loss rates and terminal wind velocities. \citet{clark18} presented NIR spectral classifications and photometry, utilising multi‑epoch spectroscopy with VLT/SINFONI and HST/WFC3 imaging to classify 88 cluster members. \citet{clark19} then updated spectral classifications for $\sim30$ per cent of those previously identified, whilst also providing new spectral classifications to take the number up to 105. Given the young age of the cluster, it is expected that insufficient time has elapsed for any massive stars to have undergone core-collapse supernovae. As such, the stars currently present are expected to be representative of the original massive stellar population of the cluster.

Radio observations (using the Very Large Array; VLA) of the Arches have been analysed by \citet{lang01}, who studied eight sources in the cluster at 5 and 8.5 GHz. This was followed by \citet{lang05} who studied 10 radio sources, showing that some of the stars originally observed in \citet{lang01} exhibit variability in their radio fluxes and presenting 22.5 GHz observations. The VLA was then used by \citet{gallego21} to study the cluster at frequencies of 6 and 10 GHz, detecting 18 radio stars. Finally, \citet{cano24} analysed 6 and 10 GHz data over multiple epochs, with 25 radio detections. The Arches has therefore been studied in the radio domain between frequencies of 5 and 22.5 GHz.

In this work we present the first Atacama Large Millimeter/submillimeter Array (ALMA) Band 3 (100 GHz; $\approx 3$ mm) and Band 6 (243 GHz; $\approx 1.3$ mm) continuum observations of the Arches cluster. We begin with a census of the sources detected in both Bands. We then determine mass-loss rates that account for the effects of wind clumping following \citet{wright75} and calculate the spectral index between the two ALMA Bands. We combine our data with that presented by \citet{lang01}, \citet{lang05} and \citet{cano24} to calculate a best-fit spectral index over the larger frequency domain now available.

A significant complication in the determination of mass-loss rates arises from the effects of wind clumping. Diagnostics that depend on the square of the density can lead to substantial overestimates of mass-loss rates if wind inhomogeneities are not properly accounted for \citep{massa03,fullerton06}. Studies have shown, however, that the magnitude of these revisions may be moderated by the effects of macroclumping, porosity and vorosity in stellar winds \citep[e.g. ][]{oskinova07,sundqvist10}. Both theoretical and observational studies indicate that clumping is already present close to the stellar surface and may vary throughout the wind \citep{runacres02,runacres05,puls06,prinja10,najarro11,sundqvist11}. Radio and millimetre observations at different frequencies probe different characteristic regions of the wind under standard assumptions for thermal free–free emission. Since millimetre observations are also less influenced by non-thermal emission, these frequencies provide a valuable means of investigating the impact of clumping on derived mass-loss rates. In this work, we use multi-frequency continuum measurements to parameterise frequency-dependent deviations from smooth-wind predictions and to explore their implications for massive-star winds in the Arches cluster.

This paper is organised as follows. Section~\ref{sec:observations} describes the observations and data reduction of the ALMA datasets. In Section~\ref{sec:analysis_results} we analyse the continuum properties of the detected sources and derive mass-loss rates and effective and frequency-dependent clumping diagnostics. Section~\ref{sec:discussion} discusses these results in the broader context of massive-star wind physics and existing theoretical and observational work. Our conclusions are presented in Section~\ref{sec:conclusions}.

\section{Observations and imaging} \label{sec:observations}
\subsection{Observations}
We use ALMA observations of the Arches cluster at Bands 3 and 6, with central frequencies of 100 and 243 GHz, respectively. Band 3 observations took place on the 19th and 23rd June 2023 (Project code 2022.1.00111.S), with a single pointing centred on the Arches cluster, covering the central 1 sq. arcmin. The array consisted of 45 antennas with baselines from 91 to 8282 m and a total on-source integration time of 4428 s ($\sim 74$ min). Band 6 observations took place on the 27th June and 10th July 2024 (Project code 2023.1.01468.S), covering the central $\sim 50$ sq. arcsec of the cluster with 6 pointings. Here, the array consisted of 42 antennas with baselines from 15 to 2516 m. The total on-source integration time per pointing was 1333 s ($\sim 22$ min). Both sets of observations were made with a total bandwidth of 8 GHz, split over 128 channels in each of four spectral windows, each with a width of 15.625 MHz. Of these, 120 channels were usable, resulting in a total effective bandwidth of 7.5 GHz. In Band 3, the spectral windows were centred on 90.5, 92.5, 102.5 and 104.5 GHz, while for Band 6 they were centred on 224, 226, 240 and 242 GHz. 

\subsection{Imaging}\label{sec:imaging}
Data calibration was performed using CASA (Common Astronomy Software Applications; v6.4.1.12 for Band 3 and v6.5.4.9 for Band 6; \citealp[]{CASA}) with the corresponding ALMA pipeline versions (v2022.2.0.68 and 2023.1.0.124, respectively). Standard a priori calibration was applied to correct for system temperature, atmospheric emission, and attenuation, followed by automated flagging of bad data. J1924--2914 was used for bandpass calibration, while J1744--3116 (Band 3) and J1752--2956 (Band 6) were used as phase calibrators.

We used the task \texttt{tclean} to implement the CASA CLEAN algorithm in order to image the data, utilising the Multi-term (Multi Scale) Multi-Frequency Synthesis (mtmfs) deconvolution \citep{rau11} with mosaicking for required for Band 6 only. For both bands, we found Briggs weighting produced the cleanest, best-resolved images. After testing a range of robust parameters, we adopted values of 1.0 (Band 3) and 0.5 (Band 6), which provided the best compromise between sensitivity and angular resolution. We further found that the default gain of 0.1 worked best for Band 3, while for Band 6 a value of 0.05 was preferred. Our final images had fitted beams of $185 \times 143$ mas and $213 \times 138$ mas, respectively, and were primary beam corrected after cleaning.

\begin{figure*}
    \centering
    \begin{subfigure}{0.49\textwidth}
        \centering
	\includegraphics[width=0.98\columnwidth]{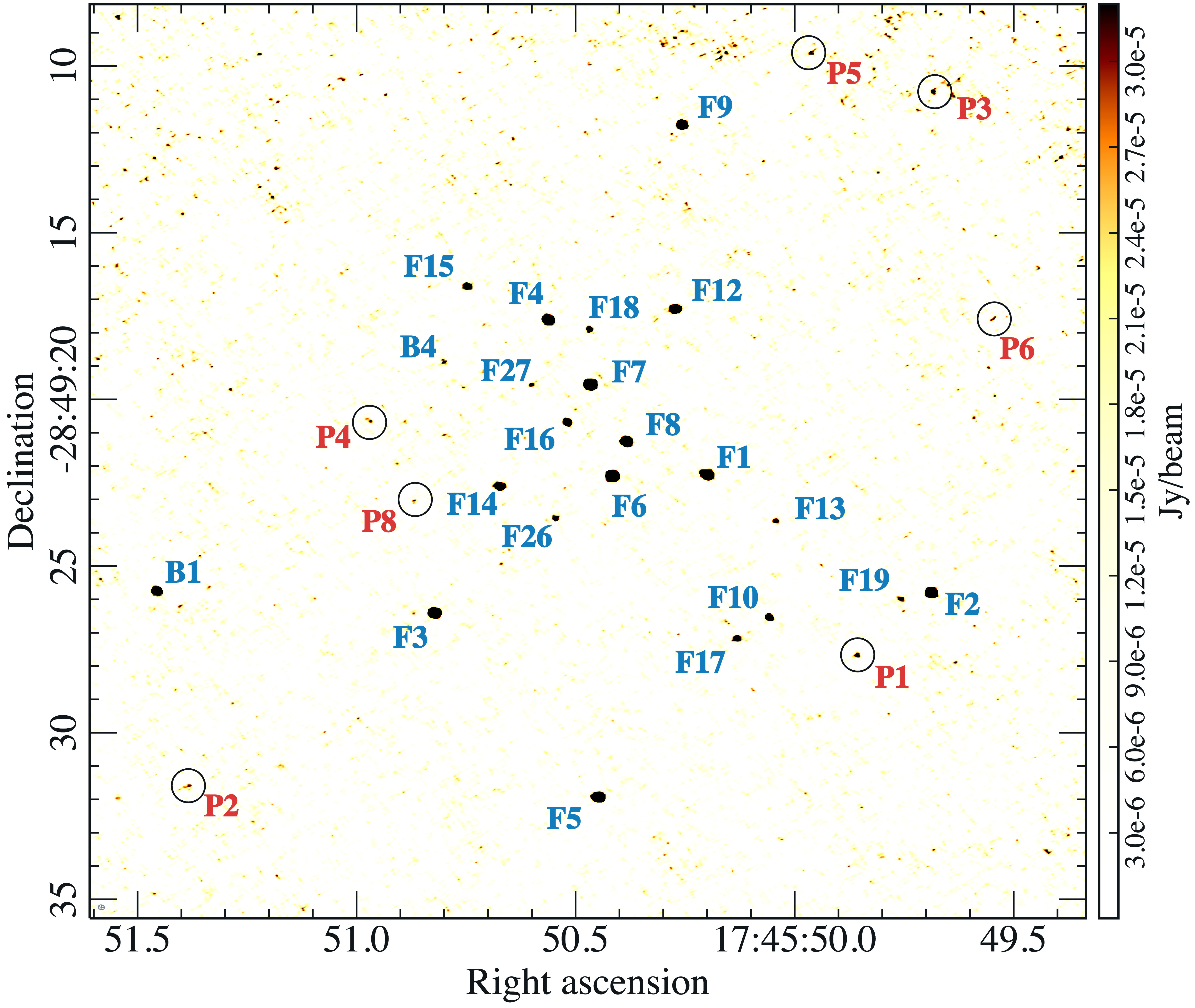}
        \caption{Band 3}
        \label{fig:b3_central}
    \end{subfigure}
    \begin{subfigure}{0.49\textwidth}
        \centering
        \includegraphics[width=0.98\columnwidth]{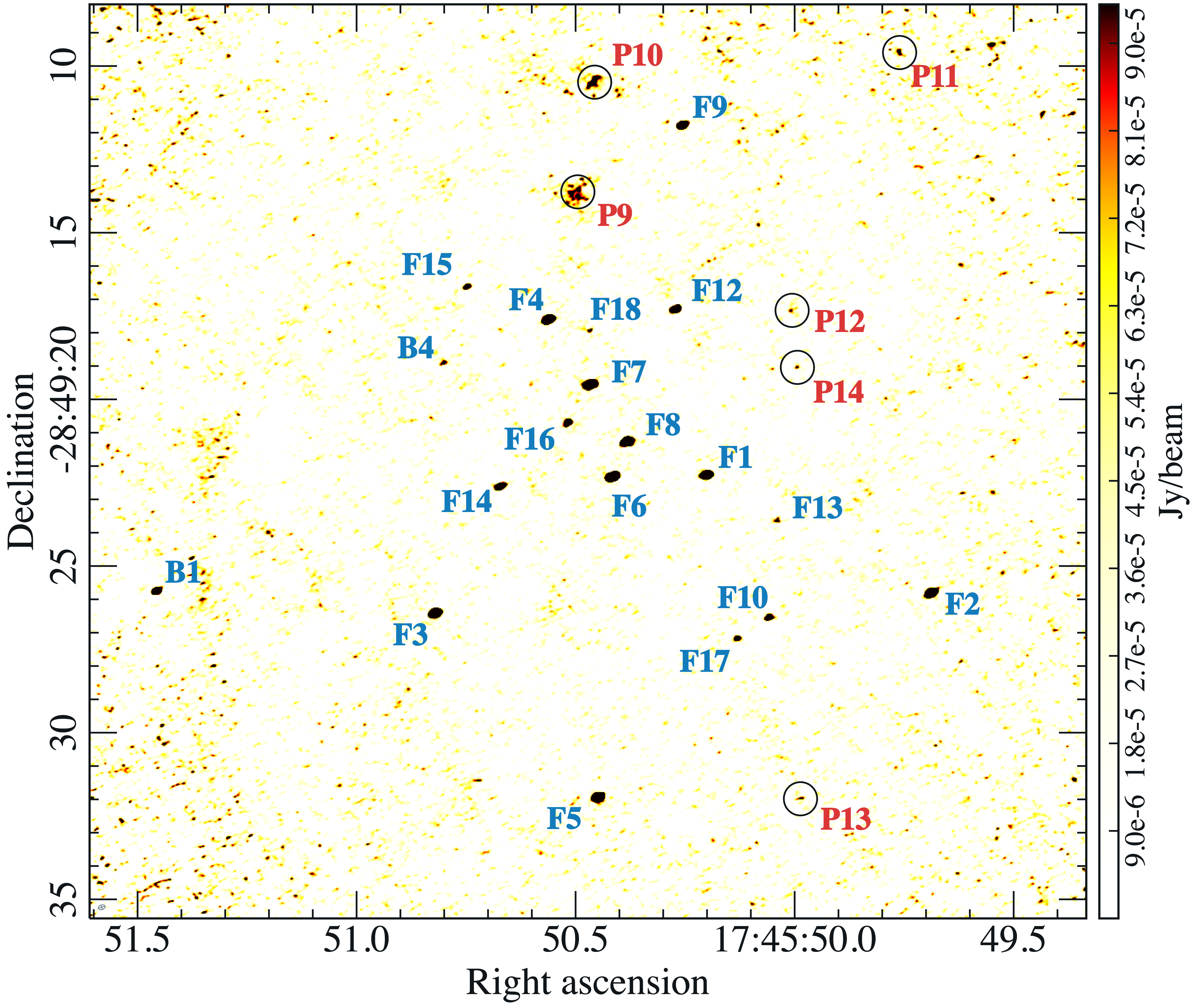}
        \caption{Band 6}
        \label{fig:b6_central}
    \end{subfigure}
    \caption{Primary-beam corrected ALMA images of the Arches cluster, labelled with stellar IDs shown in Table~\ref{tab:arches_sample} for identified sources and Table~\ref{tab:unknown_sources} for unidentified sources. The synthesised beam is shown in the lower left.}
    \label{fig:alma_band_images}
\end{figure*}

For each band, all visibilities were concatenated and two iterations of phase-only self-calibration were performed using one solution per scan. The resulting calibration tables were applied to the data prior to imaging. Flux densities were measured from the final self-calibrated continuum images.

\begin{figure}
	\includegraphics[width=1\columnwidth]{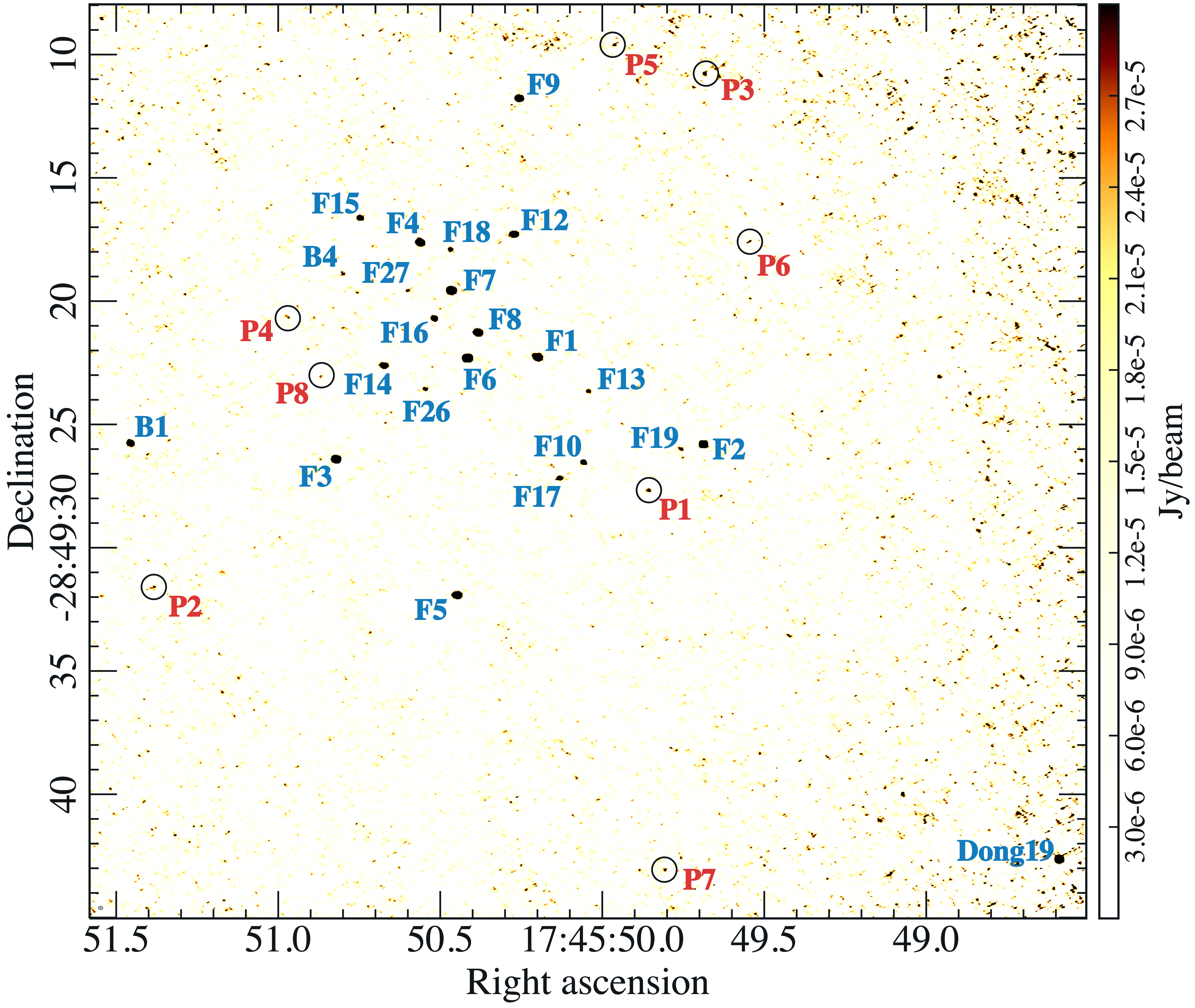}
    \caption{Wide-view primary-beam corrected ALMA Band 3 image of the Arches cluster, labelled with stellar IDs shown in Table~\ref{tab:arches_sample} for identified sources and Table~\ref{tab:unknown_sources} for unidentified sources.}
    \label{fig:b3_wide}
\end{figure}

To assess the robustness of the measured flux densities against observing block, spectral window, and imaging choices, we generated multiple independent images for each band. This included images from individual spectral windows, observing blocks, and their combinations, resulting in a total of 20 images across both bands, as well as five continuum images formed by averaging all spectral windows. We found no statistically significant differences in measured flux densities between the original deep continuum images and those produced from the alternative imaging combinations. We therefore used the secondary images for our analysis (Figs~\ref{fig:alma_band_images} and~\ref{fig:b3_wide}). The resultant flux densities for each spectral window from each observing block and the combined visibility subsets can be found in Appendix~\ref{sec:appendix:spw_fluxes}.

To check for the presence of faint stars, we used the CASA task \texttt{uvsub} to subtract model visibilities created during the cleaning process. The resultant data were then re-imaged as before. We found no additional sources within the model-subtracted image.

\subsection{Source extraction and flux density determination}
We used the Python Blob Detector and Source Finder \citep[\textsc{PyBDSF};][; version 1.12.0]{mohan15} for the detection and characterisation of sources within our images. We initially adopted the default significance/ pixel threshold of 5$\sigma$ and island boundary threshold 3$\sigma$. In subsequent runs, the significance threshold was lowered to 4.5$\sigma$ and then 4$\sigma$ in order to detect fainter sources. For each stage we imposed a hard threshold limit within \textsc{PyBDSF}.

\textsc{PyBDSF} estimates the local background and rms noise using adaptive box sizes and identifies islands of contiguous emission above the pixel threshold. An island is then grown by adding any surrounding pixels exceeding the island threshold, fitting one or more Gaussian components to each island. Source flux densities are computed as the sum of the associated Gaussian components, with positions defined by the fitted centroids. Lower thresholds were adopted only to identify candidate faint sources, with all detections subsequently verified by visual inspection.

Four sources detected in Band 3 were not detected in Band 6 above the 4$\sigma$ threshold; F19, F26, F27 and Dong19. Dong19 is not within the field-of-view of the Band 6 observations. For the remaining stars, a 3$\sigma$ upper limit on the flux density was estimated from the local rms noise at the source positions. For F27 we note that the star is visible in Fig~\ref{fig:b6_central} but fell below the detection threshold.

\subsection{Source identification}\label{sec:source_matching}
Due to the use of phase-only self-calibration, absolute astrometric information is not preserved in the final images. We instead used the positions output by \textsc{PyBDSF} as part of the Gaussian fitting. The positions of the detected sources were compared against published catalogues \citep[e.g.][]{figer02, dong11, nagata95, blum01, lang01,lang05}, primarily relying on the catalogue by \citet{clark18}. We note a systemic offset between the coordinates determined in our data and those given in \citet{clark18}. Coordinates were, on average, offset in right ascension by $+0.9^{\prime \prime}$ and in declination by $-0.5^{\prime \prime}$. The positions determined for the known sources did, however, agree with those given by \citet{cano24}. Any systematic offset did not affect source identification, as relative positions within the field are preserved and the offsets are consistent across all detected sources.

\section{Analysis and results}\label{sec:analysis_results}
\subsection{Spectral indices}
The spectral index, $\alpha$, is calculated assuming the measured flux density ($S_{\nu}$) varies as $S_{\nu} \propto \nu^{\alpha}$ for observing frequency $\nu$. Isotropic, homogeneous, thermal winds are expected to have a spectral index of $\alpha \approx 0.6$ \citep{wright75}. In colliding-wind binaries (CWBs), shocks forming exterior to the radio/mm photosphere can produce non-thermal synchrotron emission, which flattens or inverts the observed spectral index. We determined the spectral index for the stars in our sample from the complete 100 and 243 GHz flux densities in Table~\ref{tab:arches_sample} using the equation

\begin{equation} \label{eq: spectral_index}
    \alpha = \frac{\log(S_{\nu_{2}} /S_{\nu_{1}})}{\log(\nu_{2}/\nu_{1})}, \qquad \text{where} \, \nu_{1} > \nu_{2}.
\end{equation}

\noindent $S_{\nu_{1}}$ and $S_{\nu_{2}}$ are the flux densities at the observed frequencies $\nu_{1}$ and $\nu_{2}$, respectively. In this work we adopt $\nu_{1} = 243$ GHz and $\nu_{2} = 100$ GHz. The uncertainty in spectral index, $\sigma_{\alpha}$, is given by
\begin{equation} \label{eq: spectral_index_uncertainty}
    \sigma_{\alpha} = \frac{1}{\log(\nu_{2}/\nu_{1})} \times \sqrt{\left( \frac{\sigma S_{\nu_{2}}}{S_{\nu_{2}}}\right)^{2}+\left( \frac{\sigma S_{\nu_{1}}}{S_{\nu_{1}}}\right)^{2}} \,,
\end{equation}
for 100 and 243 GHz flux density uncertainties $\sigma S_{\nu_{2}}$ and $\sigma S_{\nu_{1}}$. Spectral indices are calculated using equation~(\ref{eq: spectral_index}) only for stars that are observed at both 100 GHz and 243 GHz. For stars not detected at 243 GHz, an upper limit on the spectral index is derived using the 3$\sigma$ upper limit on the Band 6 flux density. This spectral index is given in Table~\ref{tab:arches_sample}.

\begin{landscape}
\begin{figure}
	\includegraphics[width=\columnwidth]{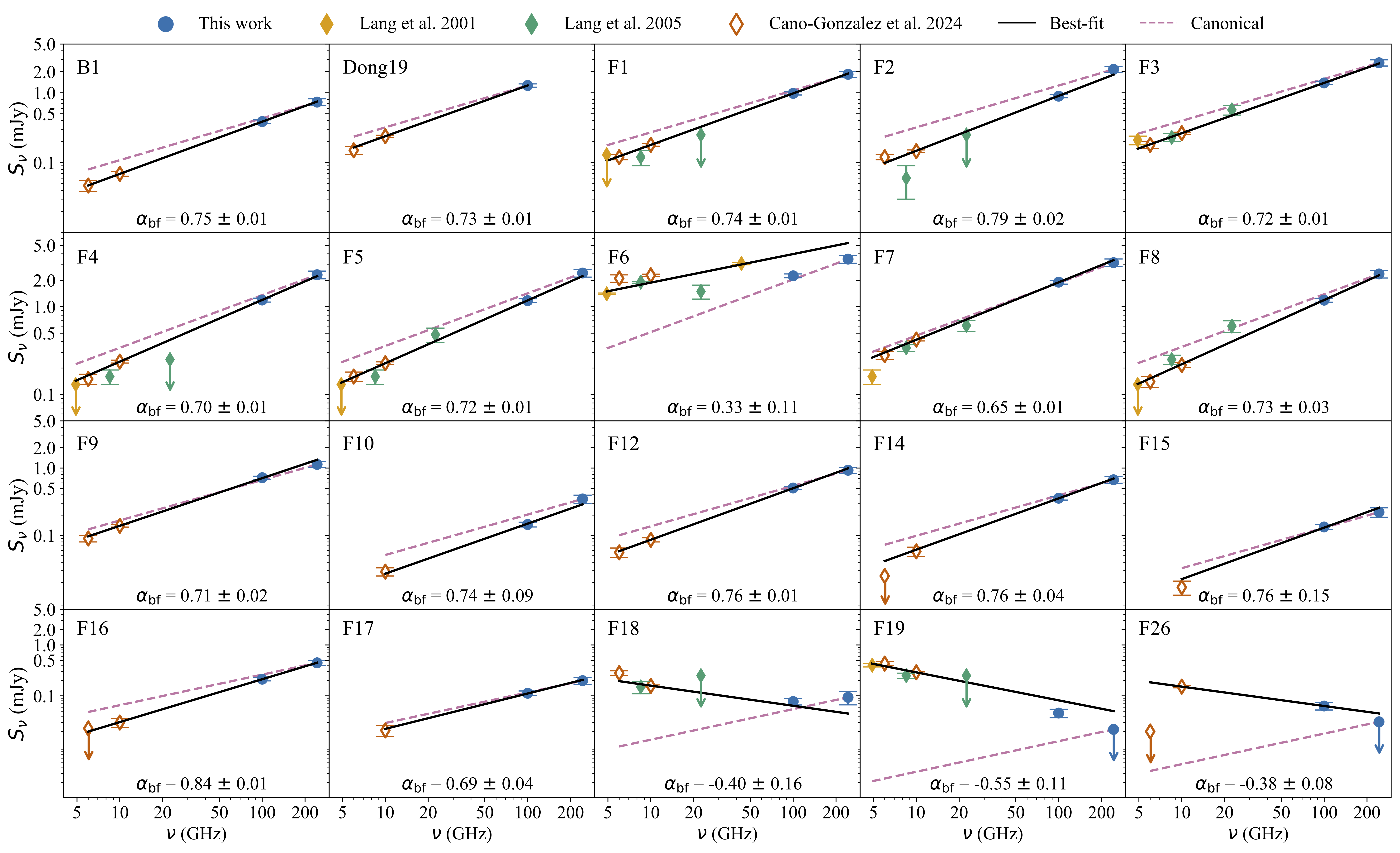}
    \caption{Flux density versus frequency plot in logarithmic space for ALMA Bands 3 and 6 observations, with VLA \textit{C}-, \textit{X}-, and \textit{K}-band observations where available. Solid lines show the best-fit spectral indices, while the dashed line indicates the canonical thermal value of $\alpha \approx 0.6$. When flux densities are upper limits, they are denoted with a downwards arrow.}
    \label{fig:canonical_plots}
\end{figure}
\end{landscape}

We further combine our data with Karl G. Jansky Very Large Array flux densities from \citet{lang01} at 4.9 GHz and a single 43.3 GHz detection of F6. We include deep \textit{C}- and \textit{X}- band fluxes (6 and 10 GHz, respectively) from \citet{cano24} and \textit{K}-band fluxes (22.5 GHz) from \citet{lang05}. Radio–mm best-fit spectral indices, $\alpha_{\text{bf}}$, were obtained using weighted least-squares fitting in log–log space, applied only to flux density detections. The weights correspond to the inverse squared uncertainties of the logarithmic flux densities, and upper limits were excluded from the fitting procedure. For several sources, the best-fit spectral index differs noticeably between weighted and unweighted fits, reflecting the uncertainties across the radio and millimetre measurements. In such cases, we adopt the weighted fit as it more appropriately accounts for the relative information content of each data point. The qualitative classification of sources based on their spectral indices is unchanged. In the case of F26 with only two frequency detections, the uncertainty on the best-fit spectral index was computed via direct error propagation using equation~(\ref{eq: spectral_index_uncertainty}). Determined values of $\alpha_{\text{bf}}$ are shown for each star in Fig.~\ref{fig:canonical_plots}.

\subsection{Mass loss}
Massive stars emit predominantly in the ultraviolet (UV) and optical parts of the electromagnetic spectrum. Emission in the mm/ radio can occur via two mechanisms: thermal free-free (Bremsstrahlung) emission from ion interactions within the wind, and non-thermal (synchrotron) emission from shocks in the colliding-wind region between CWBs. While mass-loss determination has primarily been performed spectroscopically \citep[e.g. ][]{hawcroft24}, \citet{wright75} and \citet{panagia75} showed mm/ radio measurements of flux density could be used to independently determine mass-loss rates. The mm/ radio emitting locations are in regions of the wind that have reached terminal velocity and the frequencies themselves are less affected by interstellar reddening. Mass-loss rates determined from mm/radio observations are therefore largely insensitive to interstellar reddening, which towards the GC is poorly constrained, and rely primarily on bulk wind properties, rather than detailed modelling of the photosphere.

The flux density of the observed emitted free-free radiation, $S_{\nu}$, is related to the mass-loss rate, $\dot{M}$, via

\begin{equation} \label{eq: mass_loss_rate_wright_barlow}
    S_{\nu} = 2.32\times10^{4} \left( \frac{\dot{M} \sqrt{f_{\text{cl}}}}{\mu  \varv_{\infty}} \right)^{4/3} \frac{1}{D^{2}} \left( \gamma g_{\text{ff}} \nu \overline{Z^2} \right)^{2/3} \,,
\end{equation}

\noindent where $S_{\nu}$ is measured in mJy at a frequency $\nu$ in Hz and $\dot{M}$ is measured in $M_\odot \text{ yr}^{-1}$; $ \varv_{\infty}$ is the terminal velocity (in km s$^{-1}$) and $D$ is the distance to the source (in kpc) \citep{wright75,panagia75}. $\mu$ is the mean molecular weight per ion, $Z$ the ratio of electron to ion density and $\gamma$ the mean number of electron per ion. The free-free 
Gaunt factor, $g_{\text{ff}}$, is approximated by

\begin{equation} \label{eq: gaunt_factor}
    g_{\text{ff}} \approx 9.77 \left[ 1 + 0.13 \log \left( \frac{T_{\text{e}}^{3/2}}{\nu \sqrt{\overline{Z^2}}} \right) \right] \,
\end{equation}
\noindent \citep{leitherer97}. $T_{\text{e}}$ is the electron temperature and taken to be half the effective temperature of the star, i.e. $T_{\text{e}}$ = 0.5 $T_{\text{eff}}$ \citep{drew89}.

Thermal free-free emission is sensitive to wind clumping, which can lead to overestimates of empirically derived mass-loss rates by factors of a few if wind inhomogeneities are not properly accounted for \citep{abbott81,lamers93}. We include this effect in our determination of mass-loss rates with the clumping factor, $f_{\text{cl}}$. It is assumed that the clumping factor is uniform across clumps with a value given by $f_{\text{cl}} = \langle \rho ^2 \rangle/\langle \rho \rangle^2$, the inverse of the volume filling factor\footnote{We note that some authors instead use $D_{\text{cl}}$ to represent the clumping factor, with $f$ often representing the volume filling factor.}. A smooth wind (i.e. no clumping) has $f_{\text{cl}} = 1$, whilst a value of $f_{\text{cl}} >1$ indicates a clumped wind that for a given flux density reduces the mass-loss rate. For the stars in our sample, we adopt values of $T_{\text{eff}}$ and $ \varv_{\infty}$ from \citet{martins08} for use in equations~(\ref{eq: mass_loss_rate_wright_barlow}) and~(\ref{eq: gaunt_factor}). Assuming only H and He are present in the stellar winds, the mean molecular weight per ion is given by

\begin{equation}\label{eq: He/H_to_mu}
    \mu= \frac{1\, + 4 \cdot \text{He/H}}{1\, + \text{He/H}}  \,,
\end{equation}

\noindent where we also adopt values for $\text{He/H}$ from \citet{martins08} to calculate $\mu$ in equation~(\ref{eq: He/H_to_mu}). In all calculations we adopt a distance to the Arches cluster as that of the GC at 8 kpc \citep{GRAVITY19}.

\begin{landscape}
\begin{table}
\centering
\begin{threeparttable}
\caption{ALMA flux densities of stars detected in the Arches cluster.}
\label{tab:arches_sample}
\begin{tabular}{ccccccccccccccc} % four columns, alignment for each
\hline
\hline
Source & Spectral & RA & Dec & $T_{\text{eff}}$ & $ \varv_{\infty}$ & He/H & $\mu$ & \multicolumn{2}{c}{Flux Density (mJy)} & Spectral & $\dot{M} \sqrt{f_{\text{cl}}}$ & Radio, NIR\\
\cline{9-10}
Name & Type & (J2000) & (J2000) & (K) & (km s$^{-1}$) & & & 100 GHz & 243 GHz & Index\tnote{(\textit{b})} & ($\mathrm{M}_{\odot}$ yr$^{-1}$) & Counterpart(s) \\
\hline
B1 & WN8-9h & 17 45 51.46 & -28 49 25.74 & 31700 & 1600 & 0.1 & 1.3 & 0.39 $\pm$ 0.02 & 0.74 $\pm$ 0.08 & 0.72 $\pm$ 0.33 & 2.17 $\times 10^{-5}$ & AR11, Dong79 \\
B4\tnote{(\textit{a})} & O5.5-6 Ia & 17 45 50.80 & -28 49 18.87 & 40000 & 2500 & 0.1 & 1.3 & 0.04 $\pm$ 0.01 & 0.14 $\pm$ 0.03 & 1.32 $\pm$ 0.78 & 9.52 $\times 10^{-6}$ &  \\
Dong19\tnote{(\textit{a})} & WN8-9h & 17 45 48.59 & -28 49 42.64 & 33000 & 1300 & 0.3 & 1.7 & 1.27 $\pm$ 0.07 & -- & -- & 5.13 $\times 10^{-5}$ & AR17 \\
F1 & WN8-9h & 17 45 50.20 & -28 49 22.27 & 33200 & 1400 & 0.1 & 1.3 & 0.98 $\pm$ 0.05 & 1.84 $\pm$ 0.19 & 0.71 $\pm$ 0.30 & 3.76 $\times 10^{-5}$ & AR3, Dong85 \\
F2 & WN8-9h + O5-6 Ia$^{+}$ & 17 45 49.69 & -28 49 25.80 & 33500 & 1400 & 0.3 & 1.8 & 0.90 $\pm$ 0.05 & 2.17 $\pm$ 0.22 & 1.00 $\pm$ 0.30 & 5.93 $\times 10^{-5}$ & AR10, Dong18 \\
F3 & WN8-9h & 17 45 50.82 & -28 49 26.41 & 29600 & 800 & 0.6 & 2.1 & 1.39 $\pm$ 0.07 & 2.69 $\pm$ 0.27 & 0.75 $\pm$ 0.29 & 4.84 $\times 10^{-5}$ & AR7, Dong82 \\
F4 & WN7-8h & 17 45 50.56 & -28 49 17.60 & 36800 & 1400 & 0.4 & 1.9 & 1.19 $\pm$ 0.06 & 2.31 $\pm$ 0.23 & 0.75 $\pm$ 0.29 & 6.42 $\times 10^{-5}$ & AR5, Dong81 \\
F5 & WN8-9h & 17 45 50.45 & -28 49 31.93 & 32100 & 900 & 0.8 & 2.3 & 1.17 $\pm$ 0.06 & 2.42 $\pm$ 0.25 & 0.82 $\pm$ 0.29 & 5.45 $\times 10^{-5}$ & AR8, Dong17 \\
F6 & WN8-9h & 17 45 50.42 & -28 49 22.31 & 33900 & 1400 & 0.2 & 1.5 & 2.25 $\pm$ 0.11 & 3.48 $\pm$ 0.35 & 0.49 $\pm$ 0.29 & 7.12 $\times 10^{-5}$ & AR1, Dong80 \\
F7 & WN8-9h & 17 45 50.47 & -28 49 19.56 & 32900 & 1300 & 0.3 & 1.7 & 1.90 $\pm$ 0.10 & 3.17 $\pm$ 0.32 & 0.58 $\pm$ 0.29 & 6.99 $\times 10^{-5}$ & AR4, Dong83 \\
F8 & WN8-9h & 17 45 50.38 & -28 49 21.27 & 32900 & 1000 & 1.0 & 2.5 & 1.19 $\pm$ 0.06 & 2.36 $\pm$ 0.24 & 0.77 $\pm$ 0.29 & 6.36 $\times 10^{-5}$ & AR2, Dong84 \\
F9 & WN8-9h & 17 45 50.26 & -28 49 11.77 & 36600 & 1800 & 0.1 & 1.3 & 0.72 $\pm$ 0.04 & 1.13 $\pm$ 0.12 & 0.51 $\pm$ 0.31 & 3.31 $\times 10^{-5}$ & AR15, Dong86 \\
F10 & O7-8 Ia$^{+}$ & 17 45 50.06 & -28 49 26.53 & 32200 & 1600 & 0.1 & 1.3 & 0.15 $\pm$ 0.01 & 0.35 $\pm$ 0.05 & 0.98 $\pm$ 0.42 & 1.23 $\times 10^{-5}$ &  \\
F12 & WN7-8h & 17 45 50.27 & -28 49 17.28 & 36900 & 1500 & 0.2 & 1.5 & 0.51 $\pm$ 0.03 & 0.93 $\pm$ 0.10 & 0.68 $\pm$ 0.31 & 2.80 $\times 10^{-5}$ & AR14, Dong87 \\
F13\tnote{(\textit{a})} & O7-8 Ia$^{+}$ & 17 45 50.04 & -28 49 23.65 & 35000 & 2000 & 0.1 & 1.3 & 0.06 $\pm$ 0.01 & 0.13 $\pm$ 0.03 & 0.80 $\pm$ 0.70 & 7.13 $\times 10^{-6}$ &  \\
F14 & WN8-9h & 17 45 50.67 & -28 49 22.61 & 34500 & 1400 & 0.1 & 1.3 & 0.36 $\pm$ 0.02 & 0.67 $\pm$ 0.08 & 0.71 $\pm$ 0.33 & 1.75 $\times 10^{-5}$ & AR12, Dong88 \\
F15 & O6-7 Ia$^{+}$ & 17 45 50.75 & -28 49 16.61 & 35600 & 2400 & 0.1 & 1.3 & 0.13 $\pm$ 0.01 & 0.22 $\pm$ 0.04 & 0.57 $\pm$ 0.47 & 1.30 $\times 10^{-5}$ &  \\
F16 & WN8-9h & 17 45 50.52 & -28 49 20.69 & 32200 & 1400 & 0.1 & 1.3 & 0.21 $\pm$ 0.02 & 0.45 $\pm$ 0.05 & 0.83 $\pm$ 0.37 & 1.30 $\times 10^{-5}$ & AR16 \\
F17\tnote{(\textit{a})} & O5-6 Ia$^{+}$ & 17 45 50.13 & -28 49 27.17 & 30000 & 1000 & 0.5 & 2.0 & 0.11 $\pm$ 0.01 & 0.20 $\pm$ 0.03 & 0.64 $\pm$ 0.50 & 1.01 $\times 10^{-5}$ &  \\
F18 & O4-5 Ia$^{+}$ & 17 45 50.47 & -28 49 17.90 & 36900 & 2150 & 0.1 & 1.3 & 0.08 $\pm$ 0.01 & 0.09 $\pm$ 0.03 & 0.22 $\pm$ 0.84 & 6.10 $\times 10^{-6}$ & AR9, Dong83 \\

F19\tnote{(\textit{a})} & O4-5 Ia & 17 45 49.76 & -28 49 25.98 & 40000 & 2500 & 0.1 & 1.3 & 0.05 $\pm$ 0.01 & $<0.022$ & $<-0.34$ & 6.02 $\times 10^{-6}$ & AR6 \\
F26 & O4-5 Ia & 17 45 50.55 & -28 49 23.56 & 39600 & 2600 & 0.1 & 1.3 & 0.06 $\pm$ 0.01 & $<0.031$ & $<-0.37$ & 7.97 $\times 10^{-6}$ & AR13 \\
F27\tnote{(\textit{a})} & O4-5 Ia$^{+}$ & 17 45 50.60 & -28 49 19.56 & 35000 & 2000 & 0.1 & 1.3 & 0.03 $\pm$ 0.01 & $<0.018$ & $<-0.18$ & 3.92 $\times 10^{-6}$ &  \\
\hline
\end{tabular}
\begin{tablenotes}
\item \textbf{Notes:} Standard uncertainty in right ascension and declination is $0.01^{\prime \prime}$. Mass-loss rate typical uncertainty is 0.1 dex. Nomenclature for sources from \citet{blum01}, \citet{figer02} and \citet{dong11}. Spectral types from \citet{clark18}, except Dong19 \citep{dong11}. Where available, fundamental parameters $T_{\text{eff}}$, $ \varv_{\infty}$ and He/H from \citet{martins08} with uncertainties $\pm 3000$ K, 100 km s$^{-1}$ and 30\%, respectively. Spectral index and associated uncertainty calculated using equations~(\ref{eq: spectral_index}) and~(\ref{eq: spectral_index_uncertainty}), respectively. Radio and NIR counterpart identification from \citet{lang01,lang05} and \citet{dong11}, respectively. \tnote{(\textit{a})}{Fundamental parameters determined from those of similar spectral type within the Arches.} \tnote{(\textit{b})}{Spectral index determined between ALMA frequencies only.}\\
\end{tablenotes}
\end{threeparttable}
\end{table}
\end{landscape}

\noindent We further assume that He$^+$ is the dominant ion in the mm-emitting region and adopt $Z = 1$ and $\gamma = 1$, following \citet{leitherer97}. In cases where stars are not listed in \citet{martins08}, we adopt values of $T_{\text{eff}}$, $ \varv_{\infty}$ and $\text{He/H}$ from stars of similar type within the Arches, also adopting the value of unity for both $Z$ and $\gamma$. 

Where possible we determine values of clumping-corrected mass-loss rates, $\dot{M}\sqrt{f_{\text{cl}}}$, using 243 GHz flux densities since these are expected to be less affected by non-thermal emission. When unavailable, we use the 100 GHz flux densities. All derived mass-loss rates are presented in Table~\ref{tab:arches_sample}.

\subsection{Wind clumping}
We investigate the radial behaviour of wind clumping by calculating clumping ratios, $f_{\text{cl}}^{\nu_2}/f_{\text{cl}}^{\nu_1}$ of the stars in our sample. Re-arranging equation~(\ref{eq: mass_loss_rate_wright_barlow}), we can compare the effects of clumping at two frequencies ($\nu_1$ and $\nu_2$) using the equation

\begin{equation}\label{eq: clumping_ratio}
     \frac{f_{\text{cl}}^{\nu_2}}{f_{\text{cl}}^{\nu_1}} = \frac{g_{\nu_1}}{g_{\nu_2}}\,  \frac{\nu_1}{\nu_2}  \left( \frac{S_{\nu_2}}{S_{\nu_1}}  \right)^{3/2}.
\end{equation}
\noindent Here $g_{\nu_1}$ and $g_{\nu_2}$ are the free–free Gaunt factors evaluated at frequencies $\nu_1$ and $\nu_2$, respectively. We directly compare our 100 \& 243 GHz observations with each other and with VLA data. Derived clumping factor ratios calculated using equation~(\ref{eq: clumping_ratio}) are presented in Table~\ref{tab:clump_factors}.

In a similar manner to the spectral index, we define a phenomenological clumping index, $\zeta$, assuming that $ f_{\text{cl}}^{\nu_2} / f_{\text{cl}}^{\nu_1} \propto \nu^{\zeta}$, such that between two frequencies, 
\begin{equation} \label{eq:clumping_index}
    \zeta = \frac{\log(f_{\text{cl}}^{\nu_2} / {f_{\text{cl}}^{\nu_1}})}{\log(\nu_{2}/\nu_{1})}.
\end{equation}

\begin{figure*}
	\includegraphics[width=2\columnwidth]{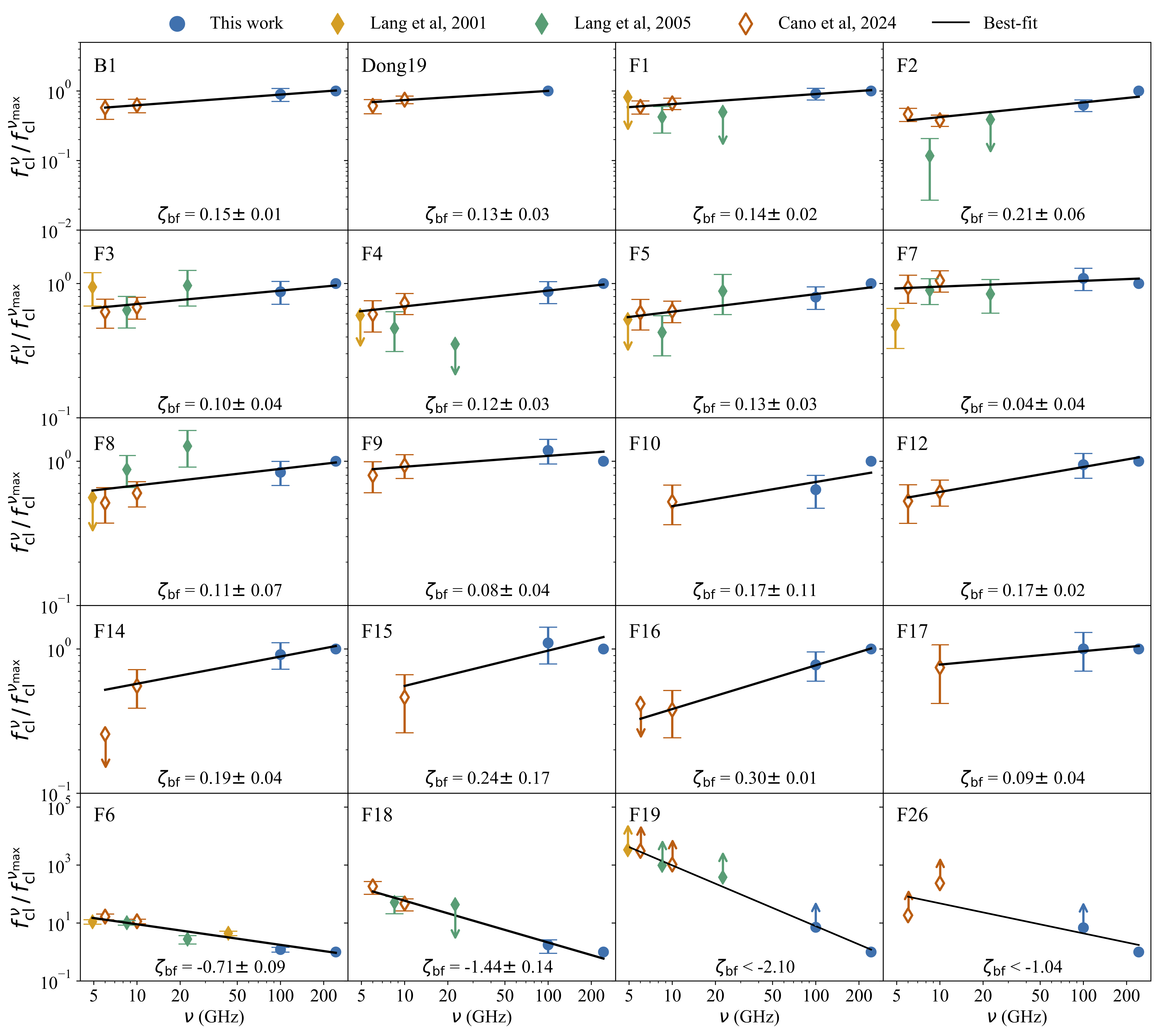}
    \caption{Clumping factor ratios vs frequency in logarithmic space for all data. Note that the value of clumping factor ratio at $\nu_\text{max}$ in each case has zero uncertainty since $f_{\text{cl}}^{\nu_\text{max}} / f_{\text{cl}}^{\nu_\text{max}}=1$ always. For F19 and F26 the value of $S_{\nu}$ at $\nu_\text{max}$ is an upper limit, making all other clumping ratios lower limits for these stars. A similar figure is shown in Appendix~\ref{sec:app_stellar_radii}, where the observation frequencies have been converted into emission radius for the thermal sources.}
    \label{fig:clumping_freq_plots}
\end{figure*}

\noindent We investigate the change in clumping by setting $\nu_1$ to $\nu_{\text{max}}$, where $\nu_{\text{max}}$ is the maximum observed frequency for a given star, and plotting the value of $ f_{\text{cl}}^{\nu} / f_{\text{cl}}^{\nu_\text{max}} $ as a function of frequency. As with spectral index, we extend equation~(\ref{eq:clumping_index}) to determine a weighted best-fit over ALMA and VLA frequencies, described by the index $\zeta_{\text{bf}}$. For stars with thermal emission, $\zeta_{\text{bf}}$ can be interpreted as tracing the radial stratification of clumping, with $\zeta_{\text{bf}} > 0$ indicating stronger clumping at smaller radii and $\zeta_{\text{bf}} < 0$ implying enhanced clumping in the outer wind. For sources with non-thermal contributions, this interpretation becomes ambiguous. At longer wavelengths the contribution from synchrotron emission is expected to increase, potentially modifying the observed ratios independently of any intrinsic change in wind structure. In such cases, $\zeta_{\text{bf}}$ serves only as a description of the frequency dependence, rather than a direct diagnostic of radial clumping. Best-fit clumping indices are shown for each star in Fig.~\ref{fig:clumping_freq_plots}.

\begin{table*}
\begin{threeparttable}
\caption{Clumping ratios}
\label{tab:clump_factors}
\centering
\begin{tabular}{cccccccc}
\hline \hline
ID  & $\left( f_{\text{cl}}^{100} / f_{\text{cl}}^{243}\right)$ & $\left( f_{\text{cl}}^{K} / f_{\text{cl}}^{100}\right)$ & $\left( f_{\text{cl}}^{X} / f_{\text{cl}}^{100}\right)$ & $\left( f_{\text{cl}}^{C} / f_{\text{cl}}^{100}\right)$ & $\left( f_{\text{cl}}^{K} / f_{\text{cl}}^{243}\right)$ & $\left( f_{\text{cl}}^{X} / f_{\text{cl}}^{243}\right)$ & $\left( f_{\text{cl}}^{C} / f_{\text{cl}}^{243}\right)$\\
\hline
B1 & 0.8 $\pm$ 0.2 & -- & 0.6 $\pm$ 0.1 & 0.5 $\pm$ 0.2 & -- & 0.5 $\pm$ 0.1 & 0.4 $\pm$ 0.1 \\
B4 & 0.4 $\pm$ 0.2 & -- & -- & -- & -- & -- & -- \\
Dong19 & -- & -- & 0.6 $\pm$ 0.1 & 0.5 $\pm$ 0.1 & -- & -- & -- \\
F1 & 0.8 $\pm$ 0.2 & 0.5 $\pm$ 0.1 & 0.6 $\pm$ 0.1 & 0.4 $\pm$ 0.1 & 0.4 $\pm$ 0.1 & 0.5 $\pm$ 0.1 & 0.3 $\pm$ 0.1 \\
F2 & 0.6 $\pm$ 0.1 & 0.5 $\pm$ 0.1 & 0.5 $\pm$ 0.1 & 0.4 $\pm$ 0.1 & 0.3 $\pm$ 0.1 & 0.3 $\pm$ 0.1 & 0.2 $\pm$ 0.1 \\
F3 & 0.8 $\pm$ 0.2 & 1.0 $\pm$ 0.3 & 0.6 $\pm$ 0.1 & 0.6 $\pm$ 0.1 & 0.7 $\pm$ 0.2 & 0.5 $\pm$ 0.1 & 0.4 $\pm$ 0.1 \\
F4 & 0.8 $\pm$ 0.1 & 0.4 $\pm$ 0.0 & 0.7 $\pm$ 0.1 & 0.5 $\pm$ 0.1 & 0.3 $\pm$ 0.0 & 0.5 $\pm$ 0.1 & 0.4 $\pm$ 0.1 \\
F5 & 0.7 $\pm$ 0.1 & 1.0 $\pm$ 0.3 & 0.6 $\pm$ 0.1 & 0.5 $\pm$ 0.1 & 0.7 $\pm$ 0.2 & 0.5 $\pm$ 0.1 & 0.4 $\pm$ 0.1 \\
F6 & 1.1 $\pm$ 0.2 & 2.0 $\pm$ 0.6 & 7.7 $\pm$ 0.9 & 12.3 $\pm$ 1.7 & 2.2 $\pm$ 0.7 & 8.4 $\pm$ 1.5 & 13.5 $\pm$ 2.5 \\
F7 & 1.0 $\pm$ 0.2 & 0.7 $\pm$ 0.2 & 0.8 $\pm$ 0.1 & 0.6 $\pm$ 0.1 & 0.7 $\pm$ 0.2 & 0.8 $\pm$ 0.1 & 0.6 $\pm$ 0.1 \\
F8 & 0.8 $\pm$ 0.1 & 1.3 $\pm$ 0.3 & 0.6 $\pm$ 0.1 & 0.4 $\pm$ 0.1 & 1.0 $\pm$ 0.3 & 0.4 $\pm$ 0.1 & 0.3 $\pm$ 0.1 \\
F9 & 1.1 $\pm$ 0.2 & -- & 0.6 $\pm$ 0.1 & 0.3 $\pm$ 0.1 & -- & 0.7 $\pm$ 0.1 & 0.3 $\pm$ 0.1 \\
F10 & 0.6 $\pm$ 0.1 & -- & 0.7 $\pm$ 0.2 & -- & -- & 0.4 $\pm$ 0.1 & -- \\
F12 & 0.9 $\pm$ 0.2 & -- & 0.5 $\pm$ 0.1 & 0.5 $\pm$ 0.1 & -- & 0.5 $\pm$ 0.1 & 0.4 $\pm$ 0.1 \\
F13 & 0.7 $\pm$ 0.3 & -- & -- & -- & -- & -- & -- \\
F14 & 0.8 $\pm$ 0.2 & -- & 0.5 $\pm$ 0.1 & 0.8 $\pm$ 0.3 & -- & 0.4 $\pm$ 0.1 & 0.7 $\pm$ 0.2 \\
F15 & 1.0 $\pm$ 0.3 & -- & 0.3 $\pm$ 0.1 & -- & -- & 0.3 $\pm$ 0.1 & -- \\
F16 & 0.7 $\pm$ 0.2 & -- & 0.4 $\pm$ 0.1 & -- & -- & 0.3 $\pm$ 0.1 & -- \\
F17 & 0.9 $\pm$ 0.3 & -- & 0.6 $\pm$ 0.2 & -- & -- & 0.5 $\pm$ 0.2 & -- \\
F18 & 1.6 $\pm$ 0.8 & 21.4 $\pm$ 4.8 & 21.9 $\pm$ 5.1 & 38.3 $\pm$ 10.2 & 34.1 $\pm$ 15.0 & 34.9 $\pm$ 15.5 & 61.0 $\pm$ 28.2 \\
F19 & 6.4 $\pm$ 1.9 & 46.5 $\pm$ 13.7 & 120.1 $\pm$ 36.0 & 301.4 $\pm$ 95.3 & 299.4 $\pm$ 22.4 & 773.1 $\pm$ 70.3 & 1940.2 $\pm$ 264.0 \\
F26 & 6.2 $\pm$ 1.6 & -- & 28.2 $\pm$ 7.6 & -- & -- & 175.5 $\pm$ 18.5 & -- \\
F27 & 5.6 $\pm$ 1.8 & -- & -- & -- & -- & -- & -- \\
\hline
\end{tabular}
\begin{tablenotes}
\item \textbf{Notes:} Clumping ratios and associated uncertainty calculated using equation~(\ref{eq: clumping_ratio}). Ratios are determined between ALMA Bands 3 and 6 (100 and 243 GHz). Where possible, ratios are also determined between each of VLA \textit{K}-, \textit{X}- and \textit{C}- bands (22.5, 10 and 6 GHz) and the individual ALMA bands.\\
\end{tablenotes}
\end{threeparttable}
\end{table*}

\section{Discussion}\label{sec:discussion}
\subsection{Mass-loss rates}
\subsubsection{Wolf-Rayets}
The clumping-scaled mass-loss rates, $\dot{M}\sqrt{f_{\text{cl}}}$, derived for the WN7-9h stars in our sample span a range of $\log(\dot{M}/\mathrm{M}_{\odot}\,\text{yr}^{-1}) \sim -4.1$ to $-4.9$, with a mean value of $\sim(4.5\pm0.2)\times10^{-5}\,\mathrm{M}_{\odot}\,\text{yr}^{-1}$. These values are consistent with previous radio studies of Galactic WR stars. For example, \citet{leitherer97} observed WR stars within 3 kpc and reported a mean mass-loss rate of $\sim4\times10^{-5}\,\mathrm{M}_{\odot}\,\text{yr}^{-1}$, while \citet{cappa04} analysed a larger Galactic sample and found a mean value of $\sim(4\pm3)\times10^{-5}\,\mathrm{M}_{\odot}\,\text{yr}^{-1}$ for classical WN stars. In particular, \citet{cano24} derived mass-loss rates for the WN7-9h stars in the Arches cluster spanning $\log(\dot{M}/\mathrm{M}_{\odot}\,\text{yr}^{-1}) \sim -4.2$ to $-5.2$\footnote{This range excludes the value derived for F6, which was based on VLA \textit{K}-band observations from \citet{lang05}, and F12, for which no mass-loss rate was determined.}, in good agreement with the range found here.

Clumping-corrected spectroscopic studies similarly find typical WN7-9h mass-loss rates of order $10^{-5}$–$10^{-4}\,\mathrm{M}_{\odot}\,\text{yr}^{-1}$ \citep{hamann06,hamann19}. Hydrodynamic wind models for luminous WNL stars predict comparable values of $\dot{M}$ for luminosities of $10^{6}$–$10^{6.3}\,\mathrm{L}_{\odot}$ \citep{grafener08}, consistent with the Arches population \citep{martins08}. Our derived rates therefore lie well within the expected range for hydrogen-rich, luminous WN stars.

The range of mass-loss rates is also comparable to that found for WN stars in Westerlund 1 \citep[Wd1; ][]{westerlund61} by \citet{fenech18}, who reported $\dot{M}\sqrt{f_{\text{cl}}}$ spanning $\log(\dot{M}/\mathrm{M}_{\odot}\,\text{yr}^{-1})\sim-4.1$ to $-4.8$. However, the mean value inferred for the Arches WN stars is higher than that reported for Wd1 ($\sim( 3.4\pm0.5)\times10^{-5}\,\mathrm{M}_{\odot}\,\text{yr}^{-1}$). This difference is unlikely to arise from methodological choices in the analysis. Differences in adopted wind terminal velocities, mean molecular weights, or distances would, if anything, reduce the discrepancy between Arches and Wd1 mass-loss rates. The weak temperature dependence of the free–free Gaunt factor implies that differences in adopted effective temperatures have a negligible effect.

Restricting the Wd1 sample to WNLh stars alone increases the mean of $\dot{M}\sqrt{f_{\text{cl}}}$ to $\sim4.8\times10^{-5}\,\mathrm{M}_{\odot}\,\text{yr}^{-1}$, comparable to that derived for the Arches. This indicates that the deviation is primarily driven by differences in sample composition. The Arches Cluster is dominated by very luminous, hydrogen-rich WNLh stars that are still undergoing core-hydrogen burning \citep{martins08, crowther10}, whereas Wd1 includes a small number of lower-luminosity WNE stars. Given the strong luminosity dependence of WR mass loss \citep{nugis00}, the inclusion of these WNE stars that exhibit lower mass-loss rates \citep[e.g. ][]{hamann19} would reduce the mean while leaving the overall range largely unaffected. Differences in wind clumping may also contribute \citep{puls06}, since radio-derived mass-loss rates scale with $\sqrt{f_{\text{cl}}}$.

\subsubsection{O super-/hypergiants}
The derived clumping-scaled mass-loss rates for the O hypergiants span $\log(\dot{M}/\mathrm{M}_{\odot}\,\text{yr}^{-1})\sim-4.9$ to $-5.4$, with a mean of ($\sim (8.8\pm0.1)\times10^{-6}\,\mathrm{M}_{\odot}\,\text{yr}^{-1}$), marginally higher than the supergiants ($\sim (7.8\pm0.1)\times10^{-6}\,\mathrm{M}_{\odot}\,\text{yr}^{-1}$). These rates are consistent with previous radio and mm constraints for evolved OB stars \citep{dougherty10,fenech18,rubio22,bjorklund23}. Since mass-loss rate increases with luminosity, our observations are inherently biased toward stars with higher $\dot{M}$. Given the small size of the hypergiant sample, comparison with Galactic counterparts is necessarily limited, serving primarily to demonstrate that the high mass-loss rates inferred here are not anomalous but consistent with stars at similar evolutionary stages. Such elevated mass-loss rates have important evolutionary implications, potentially enabling envelope stripping on timescales of $10^{4}$–$10^{5}$ yr and facilitating transitions to luminous blue variable or WR phases \citep{smith14}.

\subsubsection{Comparison with theoretical mass-loss prescriptions}
Several theoretical prescriptions are available for predicting the mass-loss rates of massive stars. For the WNLh population, commonly used prescriptions include those of \citet{grafener08} and \citet{vink17}, while for O hypergiants the prescriptions of \citet{vink01} and \citet{krticka24} are frequently adopted. These models differ in both their underlying assumptions and their intended domains of applicability. In particular, the Gr{\"a}fener prescription explicitly incorporates the role of the Eddington factor in driving dense winds, whereas the Vink prescriptions are based on Monte Carlo simulations of line-driven mass loss. Similarly, the Krti{\v{c}}ka prescription focuses on the wind initiation and acceleration region, while the Vink formulation predicts global, time-averaged mass-loss rates.

We explored the predictions of these prescriptions using representative stellar parameters from the literature. In general, the resulting mass-loss rates can differ substantially, in some cases by more than an order of magnitude depending on the adopted prescription and stellar parameters. This highlights the sensitivity of theoretical mass-loss predictions to the underlying assumptions regarding wind driving, stellar structure, and proximity to the Eddington limit.

A detailed comparison between the theoretical predictions and the mass-loss rates presented in this work is complicated by uncertainties in the fundamental stellar parameters of Arches cluster members. While the radio and mm mass-loss rates derived here are independent of stellar luminosity and radius, theoretical prescriptions require additional quantities such as luminosity, stellar mass (or surface gravity) and Eddington factor. Although effective temperatures are reasonably constrained from spectroscopic analyses, the luminosities of Arches cluster members remain subject to significant systematic uncertainties owing to the poorly constrained extinction law towards the GC. For example, \citet{najarro24} demonstrated that derived luminosities can differ by up to $\sim0.6$ dex depending on the adopted extinction prescription. Surface gravities are also poorly constrained, with values generally adopted rather than directly measured. Since luminosity and gravity together determine the Eddington factor, which plays a central role in several theoretical prescriptions, the predicted mass-loss rates are strongly dependent on assumptions that remain uncertain for the Arches population.

Furthermore, the ALMA measurements presented here constrain $\dot{M}\sqrt{f_{\rm cl}}$, whereas the theoretical prescriptions predict the underlying mass-loss rate. A direct comparison therefore requires knowledge of the radial clumping structure throughout the wind, which remains uncertain. For these reasons, we defer a detailed comparison between the observationally derived and theoretical mass-loss rates until improved constraints on both the extinction and wind clumping become available. Future low-frequency observations with next-generation facilities such as SKAO-Low \citep{bonaldi24} will provide direct constraints on the outer wind, where clumping effects are reduced, enabling more accurate determinations of the true mass-loss rates of WR stars in the GC.

\subsection{Spectral indices}
The WR stars exhibit a narrow distribution of radio-to-mm spectral indices, with the majority of WN7–9h stars clustered around $\alpha_{\text{bf}}\simeq0.7-0.8$, consistent with partially optically thick thermal free–free emission from dense stellar winds. These values are slightly steeper than the canonical $\alpha\simeq0.6$ expected for thermal outflows \citep{wright75,panagia75}, likely reflecting optical-depth effects, high wind densities, and/or clumping \citep{leitherer91}. The small dispersion in $\alpha_{\text{bf}}$ indicates broadly similar wind properties across the WR population, with any non-thermal emission strongly attenuated by free–free absorption in the inner wind. ALMA-only spectral indices (between 100 and 243 GHz) are, within uncertainties, consistent with the broadband values and cluster around $\alpha \approx 0.6-0.8$, supporting a predominantly thermal origin for the emission even at millimetre wavelengths. An exception to this is F6, which is discussed in Section~\ref{sec:f6}.

The O supergiants and hypergiants display a much broader range of spectral indices, including flat and negative values indicative of non-thermal synchrotron emission, possibly associated with wind–wind collision regions in binary systems \citep{dougherty00,debecker07}. In several cases, broadband indices are flat or negative while the ALMA-only indices remain positive, consistent with synchrotron emission dominating at cm wavelengths but becoming increasingly absorbed at mm frequencies. This behaviour suggests that O-star winds are sufficiently optically thin to allow non-thermal emission to escape at low frequencies, unlike the denser WR winds.

These findings are consistent with variability-based classifications from recent VLA studies of the Arches cluster \citep{cano24} and demonstrate the utility of combined radio–mm observations for identifying non-thermal contamination and refining radio-based mass-loss estimates.

\subsection{Radial wind structure and clumping}
Table~\ref{tab:clump_factors} shows that, for most stars, clumping affects ALMA Bands~3 and~6 at comparable levels. The main exceptions are F19, F26, and F27, which exhibit stronger inferred clumping in Band~3 (along with negative spectral indices). Comparisons with the VLA bands indicate that clumping generally has a larger effect at millimetre wavelengths than at centimetre wavelengths, with the same exceptions and the additional inclusion of F6. Particularly extreme clumping-factor ratios are inferred for F19, reaching values close to 2000 between VLA \textit{C}-band and ALMA Band~6.

The best-fit clumping indices (Fig.~\ref{fig:clumping_freq_plots}) are positive for the majority of stars, indicating increasing inferred clumping with increasing frequency. Four stars (F6, F18, F19 and F26) instead show negative clumping indices, corresponding to decreasing inferred clumping with increasing frequency. Of these, F6 is the only thermal source (see Section~\ref{sec:f6}). Since higher frequencies probe smaller radii, these results imply that, for most sources, clumping is approximately constant or decreases outward through the wind.

Wind clumping is a well-established feature of radiation-driven outflows in massive stars. Multi-wavelength diagnostics and variability studies show that smooth-wind models cannot simultaneously reproduce optical, infrared, millimetre, and radio observations, requiring structured winds at essentially all radii \citep[e.g.][]{eversberg98,lepine99,puls06}. For O-type stars, empirical studies generally find strong clumping in the inner wind that declines outward, with typical inner-wind clumping factors of $f_{\rm cl}\sim10$–50 \citep{puls06,najarro11}. Our results broadly support this picture, with most stars showing stronger clumping at smaller radii and weaker structure at larger distances. Three objects—the O hypergiant F18 and the O supergiants F19 and F26—deviate from this trend and are discussed individually in Section~\ref{sec:f18f19f26}.

Clumping is likewise ubiquitous in WR winds, as demonstrated by emission-line variability and polarimetric studies \citep{lepine99,stlouis09,davies07}. Radiative-transfer modelling typically requires inner-wind clumping factors of $f_{\rm cl}\sim4$–20 \citep{hamann98,hillier99}. More recent work suggests that clumping in WR winds generally decreases with radius, although subtype-dependent behaviour may occur \citep{chene20,sander15}. Our WR sample follows this trend in all but one case (F6), which is discussed further in Section~\ref{sec:f6}.

\subsubsection{Trends in clumping gradients}
For the subsample of stars exhibiting thermal radio-to-mm spectral indices (WNLh stars B1, Dong19, F1, F2, F3, F4, F5, F7, F8, F9, F12, F14 and F16 along with O hypergiants F10, F15 and F17), we identify a tentative empirical trend in which larger values of $\zeta_{\text{bf}}$ are associated with faster, lower-density winds. In particular, stars with higher terminal velocities tend to exhibit steeper positive clumping gradients, while objects with higher mass-loss rates show systematically flatter behaviour. Although the sample size is small and no formal statistical test is applied, this pattern is qualitatively consistent with expectations from line-driven instability theory, in which higher wind velocities promote stronger velocity perturbations and shock formation \citep[e.g.][]{owocki88}. In lower-density winds, reduced radiative damping allows shocks and associated density contrasts to persist to larger radii, producing a more pronounced radial evolution of clumping \citep{runacres02,runacres05}. Conversely, denser and slower winds are expected to homogenise more efficiently, leading to weaker clumping gradients. No clear dependence on effective temperature or luminosity is evident within this limited sample, suggesting that wind dynamics, rather than global stellar parameters, play the dominant role in setting the observed clumping behaviour. 

\subsection{Notes on individual stars}\label{sec:individual_stars}
\subsubsection{F2} 
The source F2 \citep[also designated as AR10 by][]{lang05} is a massive binary. \citet{lohr18} classed this source as an eclipsing SB2 binary after analysis of changes in radial velocity (RV). They found the system to be consistent with an $82\pm12\,\mathrm{M}_{\odot}$ WN8-9h primary and $60\pm8\,\mathrm{M}_{\odot}$ O5-6 Ia$^{+}$ secondary. Their determined orbital eccentricity of $0.075\pm0.015$ implied they were early in their evolution such that significant interaction has not occurred. The spectral indices for ALMA (1.00 $\pm$ 0.30) and for ALMA+VLA ($0.79\pm0.02$) determined in this work are both consistent with purely thermal emission. In Fig.~\ref{fig:canonical_plots} we plot the flux density of F2 as a function of frequency. We see that the thermal-like emission is not confined to higher frequencies. Notwithstanding the lower flux density at 8.5 GHz we see an overall thermal profile.

An explanation can be drawn from \citet{blomme24}. They found the CWB HD 168112 to have a spectral index that varied with orbital phase. In this case, the system has a high eccentricity ($\sim 0.75$). It was found that at apastron the system exhibits a negative spectral index, as one would expect from a CWB. However, at periastron, the spectral index is positive. The authors found that the separation between the two components at periastron was smaller than the radius of the radio photosphere and thus the non-thermal synchrotron emission was absorbed in the two winds.

To test this, we used equation 11 from \citet{wright75} to calculate the radius of the mm/cm photosphere such that at frequency $\nu$, the photospheric radius, $R(\nu)$ is given by

\begin{equation}\label{eq: photosphere_radius}
     R(\nu) = 4.0247\times10^{17}  \left( \frac{\dot{M} \sqrt{f_{\text{cl}}}}{\mu \varv_{\infty} \nu} \right)^{2/3} \frac{1}{T_{\text{eff}}^{1/2}} \left( \gamma g_{\text{ff}} \overline{Z^2} \right)^{1/3} \,\mathrm{R}_{\odot},
 \end{equation}

\noindent where we have re-scaled the radius into units of solar radius (i.e. $6.957\times10^{10}\,\text{cm}$). We use values of $\dot{M}$, $f_{\text{cl}}$, He/H, $ \varv_{\infty}$ and $T_{\text{eff}}$ from \citet{lohr18} and again assume $\gamma = Z =1$. Using equation~(\ref{eq: photosphere_radius}) we find the photosphere of the F2 secondary is within the photosphere of the primary at frequencies below $\approx35\,\text{GHz}$. Above this frequency, the secondary mm-photosphere is only partially within the photosphere of the primary, until $\approx290\,\text{GHz}$. \citet{dougherty00} noted that relativistic electrons (i.e. those producing the non-thermal emission) may be in situ if the separation of stars in a CWB are separated by a few 1000 $R_{*}$. This is not the case here, as the semi-major axis of the near-circular orbit of the F2 components is only $105\pm5\,\mathrm{R}_{\odot}$ \citep{lohr18}, or a few radii of the primary. We therefore find that the non-thermal synchrotron emission associated with the F2 binary is being absorbed by free-free absorption in the stellar wind material. We note that in their study of HD 164794, \citet{blomme14} found that large orbital separations inhibit absorption of synchrotron photons, supporting that the close proximity of the components enables synchrotron absorption.

\subsubsection{F6} \label{sec:f6}
F6 is an outlier within the WN7–9h population, exhibiting a flattened but positive broadband radio–to–mm spectral index of $\alpha_{\text{bf}}=0.33\pm0.11$. When restricted to the ALMA frequency range (100--243 GHz), the spectral index slightly steepens to $\alpha = 0.49 \pm 0.29$, consistent with partially optically thick free–free emission. This change may indicate an additional emission component contributing at cm wavelengths that is increasingly suppressed towards mm wavelengths.

Recent radio continuum surveys of the Arches cluster classify F6 as a composite source exhibiting both thermal and non-thermal characteristics, placing it among a subset of massive stars identified as candidate CWBs based on their radio properties \citep[e.g.][]{cano24}. More broadly, F6 forms part of a population of massive stars in the Arches cluster with a high inferred binary fraction, particularly among WN and O supergiant systems \citep{clark23}. Taken together, multi-wavelength constraints support an interpretation of F6 as a close, eccentric binary system in which wind--wind interactions play a significant role in shaping the observed emission properties.

Radial-velocity analysis identifies F6 as a close, pre-interaction binary with an orbital period of $13.378 \pm 0.004$~d and a high eccentricity ($e \simeq 0.6$; \citealt{clark23}). In such systems, strong shocks are expected to form at the interface between the two stellar winds, providing sites for particle acceleration and non-thermal emission, with the shock strength and geometry varying over the orbit in eccentric systems. The observed retention of a positive, though flattened, spectral index is consistent with a collision region that remains embedded within the optically thick WR wind, as expected for a relatively close binary. The relatively low-mass companion inferred for F6 likely has a weaker wind than the WR primary, displacing the wind–wind collision region closer to the secondary and embedding it more deeply within the dense WR wind, thereby enhancing free–free absorption of any non-thermal emission in the mm regime. The apparent increase in inferred wind clumping at lower frequencies may instead reflect a non-thermal contribution that enhances the cm flux. This interpretation is supported by the detection of strong 6.7 keV X-ray emission and significant emission above 7 keV from F6 \citep{wang06} and its classification as a composite (thermal + non-thermal) radio source in recent surveys of the Arches cluster \citep[e.g.][]{cano24}. 

\subsubsection{F18, F19 and F26} \label{sec:f18f19f26}
F18, F19, and F26 exhibit flat or negative radio spectral indices that distinguish them from the predominantly thermal wind sources in the sample. F18 has a best-fitting spectral index over all frequencies of $\alpha_{\text{bf}}=-0.40\pm0.16$, with an ALMA-only value of $\alpha = 0.22 \pm 0.84$, albeit with large uncertainty. F19 shows the most extreme behaviour, with a strongly negative broadband spectral index of $\alpha_{\text{bf}}=-0.55\pm0.11$ and an ALMA constraint of $\alpha < -0.34$. F26 displays a negative broadband index ($\alpha_{\text{bf}}=-0.38\pm0.08$) and an ALMA upper limit of $\alpha < -0.37$. For all three sources, the inferred wind clumping factor decreases with increasing frequency, corresponding to progressively smaller emitting radii at higher frequencies.

Such radio properties are indicative of a significant non-thermal contribution, consistent with synchrotron emission produced in shocks formed at the interface between two colliding stellar winds. Multi-epoch radio observations show variability on month-long timescales for F18 and F26, and all three stars are classified as primary CWB candidates based on their persistently negative radio spectral indices \citep{cano24}. Independent NIR radial-velocity monitoring identifies F18 as RV-variable, supporting binarity, while F19 lacks sufficient uncontaminated data for a firm RV classification due to blending with F2 \citep{clark23}. Although no RV variability was detected for F26, \citet{clark23} found the radio-derived mass-loss rate exceed that expected for its spectral type, which has been attributed to a possible non-thermal contribution, in agreement with the observed spectral indices.

The apparent increase in inferred wind clumping towards lower radio frequencies for these sources may therefore not reflect a true radial variation in wind structure, but instead arise from the increasing contribution of non-thermal emission at larger emitting radii, where free-free absorption is reduced. Shocks formed in CWBs provide a natural observational explanation for the combined spectral, variability, and flux properties of F18, F19, and F26.

\subsection{Identification and Nature of Unclassified ALMA Sources}
To assess the nature of the additional compact sources detected in the ALMA images (Table~\ref{tab:unknown_sources}), we performed a systematic cross-match against major Galactic and extragalactic catalogues using SIMBAD \citep{wenger00} and VizieR \citep{ochsenbein00}. The catalogues searched include the NVSS \citep{condon98}, FIRST \citep{becker95}, VLASS \citep{lacy20}, CORNISH \citep{hoare12}, ATLASGAL \citep{schuller09}, VVV \citep{minniti10}, and \textit{Chandra} Galactic Centre source catalogues \citep[e.g. ][]{muno09}. Matching radii between 0.5 and 2.0 arcsec were explored to evaluate the robustness of potential associations, accounting for the astrometric offset considerations discussed in Section~\ref{sec:source_matching}. 

Given the typical astrometric accuracy of ALMA for compact sources ($\lesssim 0.1$–0.2 arcsec) and the extreme source crowding in the Arches cluster, we adopt 1 arcsec as a conservative upper limit; associations appearing only at larger radii are increasingly susceptible to chance alignments. For search radii $\leq$1 arcsec, only P12 returns SIMBAD associations, both classified as ‘Unknown’ objects and lacking counterparts in any radio, sub-mm, infrared, or X-ray catalogue. Tentative positional coincidences suggest that P12 could be associated with either F79 (at a distance of 0.31 arcsec) or F113 (at 0.90 arcsec), though these identifications remain uncertain given the positional offsets and absence of multiwavelength confirmation. No sources are detected in deep radio surveys (NVSS, FIRST, VLASS, CORNISH), dust continuum emission (ATLASGAL), NIR surveys (VVV), or X-ray observations with \textit{Chandra}, arguing against an extragalactic origin or classification as compact \ion{H}{ii} regions or luminous young stellar objects. Several sources are close to the ALMA detection threshold, indicating that local noise variations and incompleteness may contribute to their apparent isolation. In the absence of robust multi-wavelength counterparts, we therefore refrain from further physical interpretation and conclude that these sources are either faint, previously unidentified cluster members or marginal detections.

\begin{table}
\centering
\caption{Unclassified ALMA sources.}
\label{tab:unknown_sources}
\setlength{\tabcolsep}{3.5pt}
\begin{tabular}{cccccc}
\hline
\hline
Source & ALMA & RA & Dec & Flux Density & Sigma \\
Name & Band & (J2000) & (J2000) & (mJy) & Threshold \\
\hline
P1 & 3 & 17 45 49.85 & -28 49 27.67 & 0.05 $\pm$ 0.01 & 5.0\\
P2 & 3 & 17 45 51.38 & -28 49 31.59 & 0.04 $\pm$ 0.01 & 4.0\\
P3 & 3 & 17 45 49.68 & -28 49 10.77 & 0.04 $\pm$ 0.01 & 4.5\\
P4 & 3 & 17 45 50.98 & -28 49 20.66 & 0.03 $\pm$ 0.01 & 4.0\\
P5 & 3 & 17 45 49.97 & -28 49 09.60 & 0.03 $\pm$ 0.01 & 4.0\\
P6 & 3 & 17 45 49.54 & -28 49 17.55 & 0.03 $\pm$ 0.01 & 4.0\\
P7 & 3 & 17 45 49.80 & -28 49 43.06 & 0.03 $\pm$ 0.01 & 4.5\\
P8 & 3 & 17 45 50.88 & -28 49 23.04 & 0.02 $\pm$ 0.01 & 4.0\\
\hline
P9 & 6 & 17 45 50.50 & -28 49 13.68 & 0.91 $\pm$ 0.07 & 4.5\\
P10 & 6 & 17 45 50.45 & -28 49 10.46 & 0.57 $\pm$ 0.06 & 4.5\\
P11 & 6 & 17 45 49.75 & -28 49 09.57 & 0.11 $\pm$ 0.03 & 4.0\\
P12 & 6 & 17 45 50.02 & -28 49 17.33 & 0.08 $\pm$ 0.02 & 4.0\\
P13 & 6 & 17 45 49.99 & -28 49 31.95 & 0.08 $\pm$ 0.02 & 4.0\\
P14 & 6 & 17 45 49.99 & -28 49 19.03 & 0.06 $\pm$ 0.02 & 4.5\\
\hline
\end{tabular}
\end{table}

\section{Conclusions}\label{sec:conclusions}
We have presented the first ALMA Band 3 (100 GHz) and Band 6 (243 GHz) continuum observations of the Arches cluster, detecting 25 massive stars and combining these measurements with archival VLA data. This broad frequency coverage enables a simultaneous assessment of mass-loss rates, emission mechanisms, and the radial structure of stellar winds in one of the most extreme star-forming environments in the Milky Way.

The derived mass-loss rates span $\log(\dot{M}/\mathrm{M}_{\odot}\,\text{yr}^{-1}) \sim -4.1$ to $-4.9$ for the WN7–9h stars and $-4.9$ to $-5.4$ for the O supergiants and hypergiants, consistent with previous radio and spectroscopic studies and with expectations for luminous, near-solar to slightly super-solar metallicity stars in the Galactic Centre. The majority of WR stars exhibit radio–mm spectral indices clustered around $\alpha=0.7$ to $0.8$, indicative of predominantly thermal free–free emission, while several O-type stars show evidence of non-thermal contributions at centimetre wavelengths, likely associated with CWBs.

The main result of this work is the characterisation of wind clumping across centimetre to millimetre wavelengths. By comparing flux densities at multiple frequencies, we quantify the radial behaviour of the clumping factor using a phenomenological clumping index, $\zeta$, to describe its frequency dependence. For most stars, clumping increases towards higher frequencies, implying stronger inhomogeneity at smaller radii and supporting theoretical predictions of structured, line-driven winds in which density contrasts are seeded near the wind base and evolve outward. The comparable clumping levels inferred between ALMA Bands 3 and 6 indicate that significant structure is already present within the millimetre photosphere. A small number of objects exhibit atypical or inverted trends, in several cases linked to binarity or non-thermal contamination. These sources highlight the importance of broadband diagnostics for disentangling intrinsic wind structure from colliding-wind contamination.

Overall, our results demonstrate that wind clumping is an important factor in shaping the emergent radio–mm emission of massive stars in the Arches cluster and must be carefully accounted for in accurate mass-loss determinations. ALMA provides a powerful probe of inner-wind structure, and future multi-frequency and time-domain observations will be essential for fully constraining the radial evolution of clumping and refining mass-loss prescriptions in extreme cluster environments.

\section*{Acknowledgements}
We thank the anonymous referee for their constructive comments and suggestions, which improved the clarity and quality of this manuscript.

This paper makes use of the following ALMA data: ADS/JAO.ALMA\#2022.1.00111.S, ADS/JAO.ALMA\#2023.1.01468.S. ALMA is a partnership of ESO (representing its member states), NSF (USA) and NINS (Japan), together with NRC (Canada), NSTC and ASIAA (Taiwan), and KASI (Republic of Korea), in cooperation with the Republic of Chile. The Joint ALMA Observatory is operated by ESO, AUI/NRAO and NAOJ.

F. Najarro gratefully acknowledges support by grant PID2022-137779OB-C41 funded by the Spanish Ministry of Science, Innovation and Universities/State Agency of Research MICIU/ AEI/10.13039/501100011033 and by “ERDF A way of making Europe” and  grant MAD4SPACE, TEC-2024/TEC-182 from Comunidad de Madrid (Spain). 

%%%%%%%%%%%%%%%%%%%%%%%%%%%%%%%%%%%%%%%%%%%%%%%%%%
\section*{Data Availability}
The ALMA data used in this paper are available in the ALMA archive at \url{https://almascience.eso.org/aq/} under project codes 2022.1.00111.S and 2023.1.01468.S.

\bibliographystyle{mnras}
\bibliography{bibliography} % if your bibtex file is called example.bib

%%%%%%%%%%%%%%%%% APPENDICES %%%%%%%%%%%%%%%%%%%%%
\appendix

\section{Spectral window flux densities} \label{sec:appendix:spw_fluxes}
Here we list the flux densities measured for each detected source in individual spectral windows and observing blocks, as well as those derived from the concatenated data. Band 3 spectral window flux densities are shown in Tables~\ref{tab:all_spw_flux_densities_b3_1}~and~\ref{tab:all_spw_flux_densities_b3_2}. Band 3 spectral window flux densities are shown in Tables~\ref{tab:all_spw_flux_densities_b6_1}~and~\ref{tab:all_spw_flux_densities_b6_2}.

%\begin{landscape}
\begin{table*}
\begin{threeparttable}
\centering
\caption{Band 3 spectral window flux densities (1).}
\label{tab:all_spw_flux_densities_b3_1}
%\footnotesize
\setlength{\tabcolsep}{4.5pt}
\begin{tabular}{ccccccccccc} % four columns, alignment for each
\hline
\hline
Source & \multicolumn{5}{c}{19th June 2023} & \multicolumn{5}{c}{23rd June 2023}\\
\cmidrule(lr){2-6}
\cmidrule(lr){7-11}
%\cline{12-16}
%\cline{17-21}
Name & 90.5 GHz & 92.5 GHz & 102.5 GHz & 104.5 GHz & Combined & 90.5 GHz & 92.5 GHz & 102.5 GHz & 104.5 GHz & Combined \\

\hline
B1 & 0.38 $\pm$ 0.04 & 0.37 $\pm$ 0.04 & 0.33 $\pm$ 0.04 & 0.43 $\pm$ 0.04 & 0.38 $\pm$ 0.03 & 0.39 $\pm$ 0.04 & 0.47 $\pm$ 0.04 & 0.41 $\pm$ 0.04 & 0.41 $\pm$ 0.04 & 0.39 $\pm$ 0.03 \\
B4 & -- & -- & -- & -- & -- & -- & -- & -- & -- & 0.05 $\pm$ 0.01 \\
Dong19 & 1.29 $\pm$ 0.09 & 1.38 $\pm$ 0.10 & 1.44 $\pm$ 0.11 & 1.30 $\pm$ 0.10 & 1.38 $\pm$ 0.08 & 1.51 $\pm$ 0.10 & 1.59 $\pm$ 0.10 & 1.28 $\pm$ 0.09 & 1.26 $\pm$ 0.08 & 1.21 $\pm$ 0.07 \\
F1 & 0.91 $\pm$ 0.06 & 1.01 $\pm$ 0.06 & 1.06 $\pm$ 0.07 & 1.05 $\pm$ 0.07 & 1.02 $\pm$ 0.06 & 0.94 $\pm$ 0.06 & 0.95 $\pm$ 0.06 & 0.94 $\pm$ 0.06 & 0.97 $\pm$ 0.06 & 0.95 $\pm$ 0.05 \\
F2 & 0.89 $\pm$ 0.06 & 0.94 $\pm$ 0.06 & 0.99 $\pm$ 0.06 & 0.96 $\pm$ 0.06 & 0.94 $\pm$ 0.05 & 0.80 $\pm$ 0.05 & 0.91 $\pm$ 0.06 & 0.87 $\pm$ 0.05 & 0.96 $\pm$ 0.06 & 0.88 $\pm$ 0.05 \\
F3 & 1.31 $\pm$ 0.07 & 1.47 $\pm$ 0.08 & 1.58 $\pm$ 0.09 & 1.55 $\pm$ 0.09 & 1.50 $\pm$ 0.08 & 1.38 $\pm$ 0.08 & 1.28 $\pm$ 0.07 & 1.33 $\pm$ 0.07 & 1.31 $\pm$ 0.07 & 1.34 $\pm$ 0.07 \\
F4 & 1.17 $\pm$ 0.07 & 1.11 $\pm$ 0.07 & 1.26 $\pm$ 0.07 & 1.39 $\pm$ 0.08 & 1.24 $\pm$ 0.07 & 1.12 $\pm$ 0.07 & 1.13 $\pm$ 0.07 & 1.13 $\pm$ 0.06 & 1.21 $\pm$ 0.07 & 1.16 $\pm$ 0.06 \\
F5 & 1.18 $\pm$ 0.07 & 1.22 $\pm$ 0.07 & 1.22 $\pm$ 0.07 & 1.28 $\pm$ 0.08 & 1.21 $\pm$ 0.06 & 1.15 $\pm$ 0.06 & 1.12 $\pm$ 0.06 & 1.20 $\pm$ 0.07 & 1.13 $\pm$ 0.07 & 1.15 $\pm$ 0.06 \\
F6 & 2.26 $\pm$ 0.12 & 2.19 $\pm$ 0.12 & 2.40 $\pm$ 0.13 & 2.57 $\pm$ 0.14 & 2.34 $\pm$ 0.12 & 2.09 $\pm$ 0.11 & 2.16 $\pm$ 0.11 & 2.21 $\pm$ 0.12 & 2.24 $\pm$ 0.12 & 2.20 $\pm$ 0.11 \\
F7 & 1.98 $\pm$ 0.11 & 1.96 $\pm$ 0.11 & 2.05 $\pm$ 0.11 & 1.99 $\pm$ 0.11 & 2.01 $\pm$ 0.10 & 1.78 $\pm$ 0.10 & 1.83 $\pm$ 0.10 & 1.84 $\pm$ 0.10 & 1.87 $\pm$ 0.10 & 1.83 $\pm$ 0.09 \\
F8 & 1.21 $\pm$ 0.07 & 1.18 $\pm$ 0.07 & 1.26 $\pm$ 0.07 & 1.32 $\pm$ 0.08 & 1.25 $\pm$ 0.07 & 1.10 $\pm$ 0.06 & 1.09 $\pm$ 0.06 & 1.18 $\pm$ 0.07 & 1.26 $\pm$ 0.07 & 1.17 $\pm$ 0.06 \\
F9 & 0.70 $\pm$ 0.05 & 0.76 $\pm$ 0.05 & 0.72 $\pm$ 0.05 & 0.76 $\pm$ 0.06 & 0.74 $\pm$ 0.04 & 0.99 $\pm$ 0.07 & 0.98 $\pm$ 0.07 & 0.75 $\pm$ 0.05 & 0.67 $\pm$ 0.05 & 0.71 $\pm$ 0.04 \\
F10 & 0.14 $\pm$ 0.02 & 0.18 $\pm$ 0.04 & 0.16 $\pm$ 0.03 & 0.17 $\pm$ 0.03 & 0.17 $\pm$ 0.02 & 0.15 $\pm$ 0.03 & 0.12 $\pm$ 0.02 & 0.11 $\pm$ 0.02 & 0.14 $\pm$ 0.02 & 0.13 $\pm$ 0.01 \\
F12 & 0.43 $\pm$ 0.04 & 0.48 $\pm$ 0.04 & 0.60 $\pm$ 0.05 & 0.46 $\pm$ 0.04 & 0.52 $\pm$ 0.03 & 0.60 $\pm$ 0.05 & 0.61 $\pm$ 0.05 & 0.58 $\pm$ 0.05 & 0.47 $\pm$ 0.04 & 0.49 $\pm$ 0.03 \\
F13 & -- & 0.07 $\pm$ 0.02 & -- & -- & 0.06 $\pm$ 0.01 & -- & -- & -- & -- & 0.05 $\pm$ 0.01 \\
F14 & 0.29 $\pm$ 0.03 & 0.31 $\pm$ 0.04 & 0.40 $\pm$ 0.04 & 0.34 $\pm$ 0.04 & 0.36 $\pm$ 0.03 & 0.48 $\pm$ 0.05 & 0.50 $\pm$ 0.04 & 0.34 $\pm$ 0.03 & 0.31 $\pm$ 0.03 & 0.36 $\pm$ 0.02 \\
F15 & -- & 0.08 $\pm$ 0.03 & 0.09 $\pm$ 0.03 & 0.11 $\pm$ 0.03 & 0.15 $\pm$ 0.02 & -- & 0.13 $\pm$ 0.03 & 0.16 $\pm$ 0.03 & 0.06 $\pm$ 0.02 & 0.11 $\pm$ 0.01 \\
F16 & 0.22 $\pm$ 0.03 & 0.25 $\pm$ 0.03 & 0.24 $\pm$ 0.03 & 0.17 $\pm$ 0.03 & 0.22 $\pm$ 0.02 & 0.13 $\pm$ 0.03 & 0.17 $\pm$ 0.02 & 0.22 $\pm$ 0.03 & 0.23 $\pm$ 0.03 & 0.20 $\pm$ 0.02 \\
F17 & 0.07 $\pm$ 0.02 & -- & -- & 0.08 $\pm$ 0.02 & 0.10 $\pm$ 0.02 & -- & -- & 0.11 $\pm$ 0.02 & 0.15 $\pm$ 0.03 & 0.10 $\pm$ 0.01 \\
F18 & -- & -- & 0.08 $\pm$ 0.03 & -- & 0.06 $\pm$ 0.01 & -- & -- & -- & 0.07 $\pm$ 0.02 & 0.08 $\pm$ 0.01 \\
F19 & -- & -- & -- & -- & -- & 0.08 $\pm$ 0.02 & -- & -- & -- & 0.05 $\pm$ 0.01 \\
F26 & -- & 0.07 $\pm$ 0.02 & -- & 0.10 $\pm$ 0.03 & 0.08 $\pm$ 0.01 & -- & -- & -- & -- & 0.04 $\pm$ 0.01 \\
F27 & -- & -- & -- & -- & -- & -- & -- & -- & -- & 0.02 $\pm$ 0.01 \\

\hline
\end{tabular}
\begin{tablenotes}
\item \textbf{Notes:} Flux densities for Arches cluster stars separated into observation date and spectral window central frequency. `Combined' data is the continuum flux density over all individual spectral windows for the observation.\\
\end{tablenotes}
\end{threeparttable}
\end{table*}
%\end{landscape}

%\begin{landscape}
\begin{table*}
\begin{threeparttable}
\centering
\caption{Band 3 spectral window flux densities (2).}
\label{tab:all_spw_flux_densities_b3_2}
\setlength{\tabcolsep}{4.5pt}
\begin{tabular}{cccccc} % four columns, alignment for each
\hline
\hline
Source & \multicolumn{5}{c}{All Observations}\\
\cmidrule(lr){2-6}
%\cline{12-16}
%\cline{17-21}
Name & 90.5 GHz & 92.5 GHz & 102.5 GHz & 104.5 GHz & Combined \\

\hline
B1 & 0.37 $\pm$ 0.03 & 0.37 $\pm$ 0.03 & 0.37 $\pm$ 0.03 & 0.43 $\pm$ 0.03 & 0.39 $\pm$ 0.02 \\
B4 & -- & -- & -- & -- & 0.04 $\pm$ 0.01 \\
Dong19 & 1.16 $\pm$ 0.07 & 1.26 $\pm$ 0.07 & 1.33 $\pm$ 0.08 & 1.28 $\pm$ 0.08 & 1.27 $\pm$ 0.07 \\
F1 & 0.94 $\pm$ 0.05 & 0.97 $\pm$ 0.05 & 0.99 $\pm$ 0.06 & 1.00 $\pm$ 0.06 & 0.98 $\pm$ 0.05 \\
F2 & 0.82 $\pm$ 0.05 & 0.92 $\pm$ 0.05 & 0.91 $\pm$ 0.05 & 0.95 $\pm$ 0.05 & 0.90 $\pm$ 0.05 \\
F3 & 1.32 $\pm$ 0.07 & 1.39 $\pm$ 0.07 & 1.43 $\pm$ 0.08 & 1.39 $\pm$ 0.07 & 1.39 $\pm$ 0.07 \\
F4 & 1.14 $\pm$ 0.06 & 1.12 $\pm$ 0.06 & 1.18 $\pm$ 0.06 & 1.29 $\pm$ 0.07 & 1.19 $\pm$ 0.06 \\
F5 & 1.17 $\pm$ 0.06 & 1.15 $\pm$ 0.06 & 1.21 $\pm$ 0.07 & 1.18 $\pm$ 0.06 & 1.17 $\pm$ 0.06 \\
F6 & 2.18 $\pm$ 0.11 & 2.19 $\pm$ 0.11 & 2.27 $\pm$ 0.12 & 2.35 $\pm$ 0.12 & 2.25 $\pm$ 0.11 \\
F7 & 1.87 $\pm$ 0.10 & 1.86 $\pm$ 0.10 & 1.92 $\pm$ 0.10 & 1.92 $\pm$ 0.10 & 1.90 $\pm$ 0.10 \\
F8 & 1.17 $\pm$ 0.06 & 1.14 $\pm$ 0.06 & 1.21 $\pm$ 0.07 & 1.25 $\pm$ 0.07 & 1.19 $\pm$ 0.06 \\
F9 & 0.72 $\pm$ 0.05 & 0.73 $\pm$ 0.04 & 0.74 $\pm$ 0.05 & 0.69 $\pm$ 0.04 & 0.72 $\pm$ 0.04 \\
F10 & 0.14 $\pm$ 0.02 & 0.16 $\pm$ 0.02 & 0.14 $\pm$ 0.02 & 0.15 $\pm$ 0.02 & 0.15 $\pm$ 0.01 \\
F12 & 0.45 $\pm$ 0.03 & 0.46 $\pm$ 0.03 & 0.60 $\pm$ 0.04 & 0.48 $\pm$ 0.03 & 0.51 $\pm$ 0.03 \\
F13 & -- & 0.05 $\pm$ 0.02 & -- & 0.07 $\pm$ 0.01 & 0.06 $\pm$ 0.01 \\
F14 & 0.35 $\pm$ 0.03 & 0.34 $\pm$ 0.03 & 0.35 $\pm$ 0.03 & 0.32 $\pm$ 0.02 & 0.36 $\pm$ 0.02 \\
F15 & 0.08 $\pm$ 0.02 & 0.11 $\pm$ 0.02 & 0.15 $\pm$ 0.02 & 0.14 $\pm$ 0.03 & 0.13 $\pm$ 0.01 \\
F16 & 0.18 $\pm$ 0.02 & 0.20 $\pm$ 0.02 & 0.24 $\pm$ 0.02 & 0.21 $\pm$ 0.02 & 0.21 $\pm$ 0.02 \\
F17 & 0.06 $\pm$ 0.01 & 0.04 $\pm$ 0.01 & 0.10 $\pm$ 0.02 & 0.13 $\pm$ 0.02 & 0.11 $\pm$ 0.01 \\
F18 & -- & 0.06 $\pm$ 0.01 & 0.06 $\pm$ 0.02 & 0.08 $\pm$ 0.02 & 0.08 $\pm$ 0.01 \\
F19 & 0.09 $\pm$ 0.02 & -- & -- & -- & 0.05 $\pm$ 0.01 \\
F26 & -- & 0.05 $\pm$ 0.01 & -- & 0.06 $\pm$ 0.02 & 0.06 $\pm$ 0.01 \\
F27 & -- & -- & -- & -- & 0.03 $\pm$ 0.01 \\

\hline
\end{tabular}
\begin{tablenotes}
\item \textbf{Notes:} Flux densities for Arches cluster stars for each spectral window over both observations. Here, the `combined' flux density represents the continuum flux density over all spectral windows and both observations.\\
\end{tablenotes}
\end{threeparttable}
\end{table*}
%\end{landscape}

\begin{table*}
\begin{threeparttable}
\centering
\caption{Band 6 spectral window flux densities (1).}
\label{tab:all_spw_flux_densities_b6_1}
\setlength{\tabcolsep}{4.5pt}
\begin{tabular}{ccccccccccc} % four columns, alignment for each
\hline
\hline
Source & \multicolumn{5}{c}{27th June 2024 (1)} & \multicolumn{5}{c}{27th June 2024 (2)}\\
\cmidrule(lr){2-6}
\cmidrule(lr){7-11}
%\cline{12-16}
%\cline{17-21}
Name & 224 GHz & 226 GHz & 240 GHz & 242 GHz & Combined & 224 GHz & 226 GHz & 240 GHz & 242 GHz & Combined \\
\hline
B1 & 0.47 $\pm$ 0.09 & 0.35 $\pm$ 0.13 & 0.50 $\pm$ 0.09 & 0.60 $\pm$ 0.13 & 0.66 $\pm$ 0.09 & 0.64 $\pm$ 0.12 & 1.05 $\pm$ 0.22 & 0.66 $\pm$ 0.13 & 0.73 $\pm$ 0.18 & 0.69 $\pm$ 0.11 \\
B4 & -- & -- & -- & -- & -- & -- & -- & -- & -- & 0.23 $\pm$ 0.05 \\
F1 & 1.73 $\pm$ 0.19 & 1.63 $\pm$ 0.21 & 1.85 $\pm$ 0.21 & 2.05 $\pm$ 0.23 & 1.81 $\pm$ 0.19 & 1.99 $\pm$ 0.22 & 2.14 $\pm$ 0.30 & 1.94 $\pm$ 0.21 & 1.89 $\pm$ 0.23 & 1.94 $\pm$ 0.21 \\
F2 & 2.02 $\pm$ 0.22 & 2.46 $\pm$ 0.30 & 2.06 $\pm$ 0.23 & 2.12 $\pm$ 0.24 & 2.15 $\pm$ 0.23 & 1.98 $\pm$ 0.23 & 4.00 $\pm$ 0.49 & 2.18 $\pm$ 0.25 & 2.31 $\pm$ 0.28 & 2.28 $\pm$ 0.24 \\
F3 & 2.44 $\pm$ 0.26 & 2.73 $\pm$ 0.31 & 2.39 $\pm$ 0.25 & 2.78 $\pm$ 0.30 & 2.64 $\pm$ 0.27 & 2.68 $\pm$ 0.30 & 7.61 $\pm$ 0.87 & 2.59 $\pm$ 0.28 & 2.69 $\pm$ 0.30 & 2.88 $\pm$ 0.30 \\
F4 & 2.00 $\pm$ 0.21 & 2.35 $\pm$ 0.28 & 2.28 $\pm$ 0.24 & 2.50 $\pm$ 0.28 & 2.37 $\pm$ 0.24 & 2.29 $\pm$ 0.25 & 4.30 $\pm$ 0.57 & 2.15 $\pm$ 0.23 & 2.29 $\pm$ 0.27 & 2.22 $\pm$ 0.24 \\
F5 & 2.11 $\pm$ 0.23 & 2.41 $\pm$ 0.30 & 2.42 $\pm$ 0.26 & 2.38 $\pm$ 0.26 & 2.38 $\pm$ 0.25 & 2.22 $\pm$ 0.24 & 3.83 $\pm$ 0.49 & 2.14 $\pm$ 0.23 & 2.74 $\pm$ 0.33 & 2.65 $\pm$ 0.28 \\
F6 & 3.08 $\pm$ 0.32 & 3.56 $\pm$ 0.39 & 3.50 $\pm$ 0.36 & 3.55 $\pm$ 0.37 & 3.46 $\pm$ 0.35 & 3.36 $\pm$ 0.36 & 6.33 $\pm$ 0.72 & 3.49 $\pm$ 0.36 & 3.98 $\pm$ 0.43 & 3.75 $\pm$ 0.38 \\
F7 & 2.84 $\pm$ 0.30 & 3.21 $\pm$ 0.36 & 3.01 $\pm$ 0.31 & 3.30 $\pm$ 0.35 & 3.16 $\pm$ 0.32 & 3.23 $\pm$ 0.34 & 9.13 $\pm$ 1.03 & 3.03 $\pm$ 0.32 & 3.38 $\pm$ 0.37 & 3.30 $\pm$ 0.34 \\
F8 & 2.11 $\pm$ 0.23 & 2.64 $\pm$ 0.31 & 2.24 $\pm$ 0.24 & 2.22 $\pm$ 0.25 & 2.34 $\pm$ 0.24 & 2.37 $\pm$ 0.27 & 3.64 $\pm$ 0.47 & 2.52 $\pm$ 0.28 & 2.34 $\pm$ 0.27 & 2.25 $\pm$ 0.24 \\
F9 & 1.17 $\pm$ 0.15 & 1.07 $\pm$ 0.17 & 1.15 $\pm$ 0.15 & 0.96 $\pm$ 0.16 & 1.08 $\pm$ 0.13 & 1.12 $\pm$ 0.15 & 1.39 $\pm$ 0.26 & 1.06 $\pm$ 0.14 & 0.85 $\pm$ 0.17 & 1.09 $\pm$ 0.13 \\
F10 & 0.21 $\pm$ 0.05 & -- & 0.17 $\pm$ 0.05 & -- & 0.28 $\pm$ 0.05 & 0.19 $\pm$ 0.05 & -- & 0.46 $\pm$ 0.09 & -- & 0.35 $\pm$ 0.07 \\
F12 & 0.80 $\pm$ 0.10 & 0.91 $\pm$ 0.15 & 0.77 $\pm$ 0.11 & 0.92 $\pm$ 0.13 & 0.93 $\pm$ 0.11 & 0.99 $\pm$ 0.13 & 1.16 $\pm$ 0.22 & 0.79 $\pm$ 0.11 & 1.11 $\pm$ 0.17 & 1.02 $\pm$ 0.12 \\
F13 & -- & -- & -- & -- & -- & -- & -- & -- & -- & -- \\
F14 & 0.49 $\pm$ 0.08 & 0.28 $\pm$ 0.08 & 0.79 $\pm$ 0.11 & 0.64 $\pm$ 0.12 & 0.57 $\pm$ 0.07 & 0.84 $\pm$ 0.12 & 0.49 $\pm$ 0.16 & 0.62 $\pm$ 0.11 & 0.63 $\pm$ 0.14 & 0.65 $\pm$ 0.09 \\
F15 & -- & -- & -- & -- & 0.27 $\pm$ 0.05 & -- & -- & -- & -- & 0.14 $\pm$ 0.04 \\
F16 & 0.48 $\pm$ 0.08 & -- & 0.42 $\pm$ 0.09 & 0.34 $\pm$ 0.09 & 0.39 $\pm$ 0.06 & 0.47 $\pm$ 0.09 & -- & 0.30 $\pm$ 0.07 & 0.51 $\pm$ 0.11 & 0.49 $\pm$ 0.07 \\
F17 & -- & -- & -- & -- & 0.22 $\pm$ 0.05 & -- & -- & -- & -- & -- \\
F18 & -- & -- & -- & -- & -- & -- & -- & -- & -- & 0.12 $\pm$ 0.05 \\

\hline
\end{tabular}
\begin{tablenotes}
\item \textbf{Notes:} Flux densities for Arches cluster stars separated into observation date and spectral window central frequency for the first and second observations in Band 6. `Combined' data is the continuum flux density over all individual spectral windows for each observation.\\
\end{tablenotes}
\end{threeparttable}
\end{table*}

\begin{table*}
\begin{threeparttable}
\centering
\caption{Band 6 spectral window flux densities (2).}
\label{tab:all_spw_flux_densities_b6_2}
\setlength{\tabcolsep}{4.5pt}
\begin{tabular}{ccccccccccc} % four columns, alignment for each
\hline
\hline
Source & \multicolumn{5}{c}{10th July 2024} & \multicolumn{5}{c}{All Observations}\\
\cmidrule(lr){2-6}
\cmidrule(lr){7-11}
%\cline{12-16}
%\cline{17-21}
Name & 224 GHz & 226 GHz & 240 GHz & 242 GHz & Combined & 224 GHz & 226 GHz & 240 GHz & 242 GHz & Combined \\

\hline
B1 & 0.91 $\pm$ 0.14 & 0.45 $\pm$ 0.11 & 0.62 $\pm$ 0.09 & 0.69 $\pm$ 0.13 & 0.78 $\pm$ 0.10 & 0.67 $\pm$ 0.09 & 0.73 $\pm$ 0.12 & 0.58 $\pm$ 0.08 & 0.76 $\pm$ 0.11 & 0.74 $\pm$ 0.08 \\
B4 & -- & -- & -- & -- & -- & -- & -- & -- & -- & 0.14 $\pm$ 0.03 \\
F1 & 1.64 $\pm$ 0.18 & 1.97 $\pm$ 0.23 & 1.65 $\pm$ 0.18 & 1.97 $\pm$ 0.22 & 1.83 $\pm$ 0.19 & 1.75 $\pm$ 0.18 & 1.83 $\pm$ 0.20 & 1.76 $\pm$ 0.18 & 1.95 $\pm$ 0.21 & 1.84 $\pm$ 0.19 \\
F2 & 0.91 $\pm$ 0.14 & 2.31 $\pm$ 0.28 & 1.88 $\pm$ 0.21 & 2.34 $\pm$ 0.26 & 2.06 $\pm$ 0.21 & 1.96 $\pm$ 0.20 & 2.30 $\pm$ 0.25 & 2.07 $\pm$ 0.22 & 2.26 $\pm$ 0.24 & 2.17 $\pm$ 0.22 \\
F3 & 2.49 $\pm$ 0.26 & 2.67 $\pm$ 0.29 & 2.52 $\pm$ 0.27 & 2.62 $\pm$ 0.28 & 2.59 $\pm$ 0.26 & 2.62 $\pm$ 0.27 & 2.78 $\pm$ 0.29 & 2.48 $\pm$ 0.25 & 2.74 $\pm$ 0.29 & 2.69 $\pm$ 0.27 \\
F4 & 2.19 $\pm$ 0.23 & 2.54 $\pm$ 0.28 & 2.22 $\pm$ 0.24 & 2.45 $\pm$ 0.26 & 2.43 $\pm$ 0.25 & 2.16 $\pm$ 0.22 & 2.44 $\pm$ 0.26 & 2.24 $\pm$ 0.23 & 2.40 $\pm$ 0.25 & 2.31 $\pm$ 0.23 \\
F5 & 2.15 $\pm$ 0.23 & 2.91 $\pm$ 0.34 & 2.09 $\pm$ 0.23 & 2.21 $\pm$ 0.25 & 2.26 $\pm$ 0.23 & 2.14 $\pm$ 0.22 & 2.73 $\pm$ 0.29 & 2.27 $\pm$ 0.24 & 2.44 $\pm$ 0.26 & 2.42 $\pm$ 0.25 \\
F6 & 3.17 $\pm$ 0.33 & 3.42 $\pm$ 0.36 & 3.33 $\pm$ 0.34 & 3.56 $\pm$ 0.37 & 3.33 $\pm$ 0.34 & 3.22 $\pm$ 0.33 & 3.49 $\pm$ 0.36 & 3.45 $\pm$ 0.35 & 3.66 $\pm$ 0.37 & 3.48 $\pm$ 0.35 \\
F7 & 2.90 $\pm$ 0.30 & 3.24 $\pm$ 0.35 & 2.84 $\pm$ 0.30 & 3.04 $\pm$ 0.32 & 3.04 $\pm$ 0.31 & 2.95 $\pm$ 0.30 & 3.37 $\pm$ 0.35 & 2.97 $\pm$ 0.30 & 3.24 $\pm$ 0.33 & 3.17 $\pm$ 0.32 \\
F8 & 1.98 $\pm$ 0.21 & 2.72 $\pm$ 0.29 & 2.19 $\pm$ 0.23 & 2.43 $\pm$ 0.26 & 2.32 $\pm$ 0.24 & 2.26 $\pm$ 0.24 & 2.46 $\pm$ 0.26 & 2.37 $\pm$ 0.24 & 2.40 $\pm$ 0.25 & 2.36 $\pm$ 0.24 \\
F9 & 0.94 $\pm$ 0.12 & 1.05 $\pm$ 0.16 & 1.04 $\pm$ 0.13 & 1.25 $\pm$ 0.16 & 1.08 $\pm$ 0.12 & 1.09 $\pm$ 0.12 & 1.08 $\pm$ 0.14 & 1.16 $\pm$ 0.13 & 1.16 $\pm$ 0.14 & 1.13 $\pm$ 0.12 \\
F10 & 0.22 $\pm$ 0.06 & -- & 0.36 $\pm$ 0.07 & 0.26 $\pm$ 0.08 & 0.32 $\pm$ 0.05 & 0.26 $\pm$ 0.05 & 0.28 $\pm$ 0.07 & 0.37 $\pm$ 0.06 & 0.25 $\pm$ 0.05 & 0.35 $\pm$ 0.05 \\
F12 & 0.73 $\pm$ 0.10 & 1.12 $\pm$ 0.15 & 0.85 $\pm$ 0.11 & 0.83 $\pm$ 0.12 & 0.88 $\pm$ 0.10 & 0.82 $\pm$ 0.10 & 0.96 $\pm$ 0.12 & 0.83 $\pm$ 0.09 & 1.00 $\pm$ 0.12 & 0.93 $\pm$ 0.10 \\
F13 & -- & -- & -- & -- & 0.08 $\pm$ 0.03 & -- & -- & 0.10 $\pm$ 0.03 & -- & 0.13 $\pm$ 0.03 \\
F14 & 0.51 $\pm$ 0.08 & 0.44 $\pm$ 0.09 & 0.80 $\pm$ 0.11 & 0.50 $\pm$ 0.09 & 0.61 $\pm$ 0.08 & 0.60 $\pm$ 0.07 & 0.53 $\pm$ 0.09 & 0.71 $\pm$ 0.09 & 0.65 $\pm$ 0.10 & 0.67 $\pm$ 0.08 \\
F15 & 0.24 $\pm$ 0.05 & -- & -- & -- & 0.24 $\pm$ 0.04 & 0.20 $\pm$ 0.04 & -- & 0.21 $\pm$ 0.04 & -- & 0.22 $\pm$ 0.04 \\
F16 & 0.28 $\pm$ 0.06 & -- & 0.41 $\pm$ 0.06 & 0.24 $\pm$ 0.06 & 0.36 $\pm$ 0.05 & 0.56 $\pm$ 0.09 & 0.51 $\pm$ 0.09 & 0.43 $\pm$ 0.06 & 0.46 $\pm$ 0.08 & 0.45 $\pm$ 0.05 \\
F17 & -- & -- & -- & -- & 0.18 $\pm$ 0.04 & -- & 0.19 $\pm$ 0.05 & -- & 0.25 $\pm$ 0.05 & 0.20 $\pm$ 0.03 \\
F18 & 1.78 $\pm$ 0.19 & -- & -- & -- & -- & -- & -- & -- & -- & 0.09 $\pm$ 0.03 \\
\hline
\end{tabular}
\begin{tablenotes}
\item \textbf{Notes:} Flux densities for Arches cluster stars separated into observation date and spectral window central frequency for the final observation in Band 6, along with `Combined' data is the continuum flux density over all individual spectral windows for the observation. Also shown are flux densities for Arches cluster stars for each spectral window over both observations and the `combined' flux density of the continuum over all spectral windows and both observations.\\
\end{tablenotes}
\end{threeparttable}
\end{table*}

\section{Clumping index as a function of emission radius} \label{sec:app_stellar_radii}

\begin{figure*}
	\includegraphics[width=2\columnwidth]{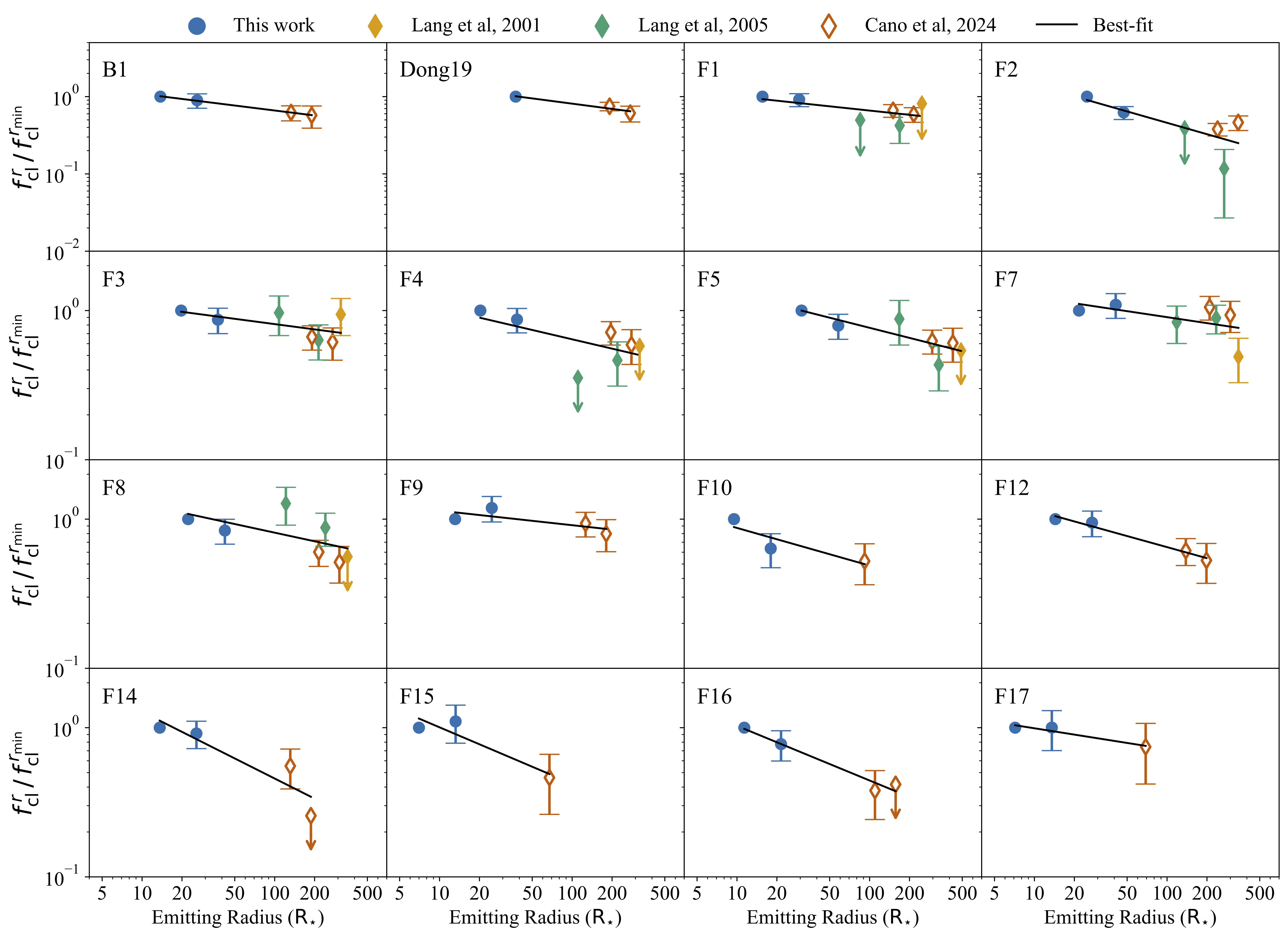}
    \caption{Clumping factor ratio as a function of characteristic emission radius, $R(\nu)$, expressed in units of stellar radius \citep[from][]{martins08}, for the thermal sources in our sample. The emission radius at each frequency is calculated following equation 11 of \citet{wright75}. Since F6 has a non-thermal component, it is not included here.}
    \label{fig:clumping_stellar_plots}
\end{figure*}

For thermal sources, the observing frequency can be converted to a characteristic emission radius, $R(\nu)$, using equation 11 of \citet{wright75}, such that the measured flux density is assumed to arise from a (mm/ radio) photosphere of radius $R$ cm. We adopt values of $\dot{M}\sqrt{f_{\text{cl}}}$, He/H, $\varv_{\infty}$ and $T_{\text{eff}}$ from Table~\ref{tab:arches_sample}, assuming $\gamma = Z = 1$. The resulting radii are expressed in units of stellar radius using the values given by \citet{martins08}. In Fig.~\ref{fig:clumping_stellar_plots} we present the clumping factor ratio as a function of emission radius, illustrating the physical interpretation of the frequency-dependent analysis.

%%%%%%%%%%%%%%%%%%%%%%%%%%%%%%%%%%%%%%%%%%%%%%%%%%

% Don't change these lines
\bsp	% typesetting comment
\label{lastpage}
\end{document}